\documentclass{aa}

\usepackage{graphicx}
\usepackage{txfonts}
\usepackage{placeins}

\usepackage{hyperref}
\usepackage{float}
\usepackage{color}

\begin{document}
\bibpunct{(}{)}{;}{a}{}{,} % A&A style 

    %\title{Planet formation in wind-driven discs via pebble accretion}
    %\title{Planet formation in wind-driven accretion discs via pebble accretion}
    %\title{Fast-wind discs generate peas-in-a-pod planetary systems}
    \title{Similar-mass \textcolor{black}{versus} diverse-mass planetary systems in wind-driven accretion discs}
    \author{Yunpeng Zhao\inst{1} \and Soko Matsumura\inst{2}}
    \institute{University of Dundee, School of Science and Engineering, Dundee DD1 4HN, UK \email{2440931@dundee.ac.uk}
    \and 
    University of Dundee, School of Science and Engineering, Dundee DD1 4HN, UK
    \email{s.matsumura@dundee.ac.uk}}
    \date{Received; accepted}

% \abstract{}{}{}{}{} 
% 5 {} token are mandatory
 
\abstract
{Many close-in multiple-planet systems show a peas-in-a-pod trend,
where sizes, masses, and orbital spacing of neighbouring planets are comparable to each other. 
On the other hand, some planetary systems have a more diverse \textcolor{black}{size and mass} distribution, including the Solar System. Classical planet formation
models tend to produce the former type of planetary systems rather
than the latter, and \textcolor{black}{the origin of} their difference is not well understood.}
{Recent observational and numerical studies support the disc evolution
that is \textcolor{black}{largely} driven by magnetic winds rather than by the traditional disc's
viscosity alone. In such a wind-driven accretion disc, the disc mass accretion
rate varies radially, \textcolor{black}{instead of being constant in radius as in the classical viscously accreting disc.}
We investigate how the wind's efficiency in removing the disc mass 
affects \textcolor{black}{the outcome of planet formation and migration.}}
%affects formation and migration of planets and shed light on the diversity of planetary systems.}
%{Planet formation in wind-driven discs is not well-studied. We numerically investigated planet formation in such discs, examining how initial conditions such as disc mass, \(\alpha_{\text{total}}\), metallicity, and the magnetic lever arm \(\lambda\) affect the outcomes of planet formation.}
{We \textcolor{black}{performed} single-core planet formation simulations via pebble accretion
in wind-driven accretion discs. We \textcolor{black}{varied} the wind's efficiency via
the magnetic lever arm parameter $\lambda$ and \textcolor{black}{studied} the outcome
of planet formation and migration by considering a range of initial disc masses and \textcolor{black}{disc accretion timescales}.}
%{We performed single-core formation of planets via the pebble accretion model in wind-driven discs. We varied different initial conditions of the model and quantitatively analyzed the differences.}
{Our simulations \textcolor{black}{show} that higher $\lambda$ discs with less wind mass loss lead to 
faster formation and migration of planets and tend to generate similar-mass planetary systems,
while lower $\lambda$ discs lead to slower formation and migration as well as more diverse-mass planetary systems.
%These two-types of systems are approximately separated at $\lambda\sim2$. 
Furthermore, we find that \textcolor{black}{a planetary system with a mass jump happens for all $\lambda$ cases as long as the planet formation timescale is comparable to the disc accretion timescale, but the jump is larger for lower $\lambda$ discs.}
The super-Earth systems accompanied by cold Jupiters \textcolor{black}{can be generated in such systems, and we find their frequencies are} higher in metal-rich discs, \textcolor{black}{which agrees} with the observational trend. 
Our simulations indicate that similar-mass and diverse-mass systems are approximately separated at $\lambda\sim2-3$.

%
%
%Furthermore, metallicity impacts the mass and orbital radii of planets,
%where higher metallicity leads to heavier planetary cores and larger
%orbital radii for giant planets. 
%
% Finally, we have also compared the adopted disc model with three different star forming regions and
% found that the disc model reproduces the evolution of slopes of the $\log(M_D)-\log(\dot{M}_*)$ distributions well 
% but predicted mass accretion rates are lower by 1-2 orders of magnitude.
}
%with higher metallicity leading to heavier planetary cores and larger
%orbital radii for giant planets. 
%Finally, the initial disc mass $M_{D,0}$
%significantly affects the occurrence rate of giant planets, with higher
%$M_{D,0}$ and $\alpha_{\text{total}}$ promoting giant planet formation.}
%{Our findings indicate that planetary cores and maximum planetary masses are generally lower in wind-driven discs due to lower \(\alpha_{\text{ss}}\) and pebble isolation mass. The magnetic lever arm \(\lambda\) influences the mass diversity of planetary systems, with smaller \(\lambda\) leading to closely paired planets with significant mass differences, while larger \(\lambda\) results in widely separated pairs with similar masses. Wind-driven discs also enhance the survival rate of giant planets by slowing type II migration. Furthermore, metallicity impacts the mass and orbital radii of planets, with higher metallicity leading to heavier planetary cores and larger orbital radii for giant planets. Finally, the initial disc mass \(M_{D,0}\) significantly affects the occurrence rate of giant planets, with higher \(M_{D,0}\) and \(\alpha_{\text{total}}\) promoting giant planet formation.}
%{conclusions}


\maketitle

\section{Introduction}
\label{sec:Introduction}

Exoplanets in multiple-planet systems are often described by the term \textcolor{black}{`peas in a pod'}, 
because they have \textcolor{black}{a} similar radius and/or mass and their orbits are evenly spaced \citep[e.g.][]{Millholland2017,Weiss2018,Rosenthal2024}.
However, the extent to which such trends apply \textcolor{black}{has been} put into question lately. 
Recent studies have shown that the similarity trend is stronger for radii than for masses \citep[e.g.,][]{Otegi2022,Mamonova2024}, 
while the orbital period spacing correlation may or may not exist \citep{Weiss2018,Mamonova2024}. 
%
%Furthermore, the similarity trends may not hold for certain groups of planets.
%Furthermore, such similarity trends may only hold for neighbouring planets that are separated relatively closely from each other.  
Regarding this latter point, \cite{Jiang2020} studied the orbital spacing of Kepler multiple-planet systems with four or more planets and 
found that the transition from correlation to non-correlation of period ratios of adjacent planets may abruptly occur 
for relatively closely separated planets with the period ratio of $\sim1.5-1.7$. 
%Thus, the similarity in orbital spacing may only hold for neighbouring planets that are separated relatively closely from each other.

Furthermore, since these various similarity trends are determined generally for close-in planets, 
it is unclear whether such a trend is applicable to planets far from the central star.
\cite{Millholland2022} evaluated the detectability of hypothetical planets beyond the observed transiting planets 
by considering the similarity trends of radius and period ratio and 
proposed that the compact multiple-planet systems may be isolated or the peas-in-a-pod trend may break down in the outer region ($\sim100-300\,$days).
On the other hand, \cite{Rosenthal2024} studied RV-detected systems with multiple giant planets over a wide range of 
orbital radii (between 0.1 and 10\,au) and found a mass similarity in the same system. 

Finally, these similarity trends appear to be weaker for giant planets with $\gtrsim10\,R_{\oplus}$ and $\gtrsim100\,M_{\oplus}$ \citep[e.g.,][]{Mamonova2024}.
This may not be too surprising since systems with multiple giant planets often undergo dynamical instabilities involving (drastic) changes of the orbits and the number of 
planets \citep[e.g.,][]{Chatterjee2008,Juric2008}, while the compact multiple-planet systems are more likely to have been evolving quiescently.
There are also planetary systems involving giant planets that show a diversity in planetary masses as \textcolor{black}{that} in the Solar System rather than a similarity. 
Warm Jupiters are known to be less isolated compared to hot Jupiters and are often accompanied by low-mass planets \citep[e.g.,][]{Steffen2012,Huang2016}. 
Also, a positive correlation has been observed for inner super Earths and outer cold Jupiters, especially around metal-rich stars \citep{Zhu2018,Bryan2019,Bryan2024}.
Considering \textcolor{black}{all of these observations}, it is important to understand what determines the observed similarity trends and to what extent such trends hold.

%Furthermore, since these trends are determined generally for close-in planets, 
%it is unclear whether such a trend is applicable to planets far from the central star.
%\cite{Millholland2022} evaluated the detectability of hypothetical planets beyond the observed transiting planets and 
%proposed that the compact multiple-planet systems may be isolated or the peas-in-a-pod trend may break down in the outer region ($\sim100-300\,$days).
%On the other hand, for systems with multiple giant planets between 0.1 and 10\,au, \cite{Rosenthal2023} 
%found a mass similarity in the same system. 
%
%While some multiple-planet systems show a peas-in-a-pod trend, others including the Solar System don't follow such a trend.
%For example, warm Jupiters are known to be less isolated compared to hot Jupiters and are often accompanied by low-mass planets \citep[e.g.,][]{Steffen2012,Huang2016}. 
%Also, a positive correlation has been observed for inner super Earths and outer cold Jupiters, especially around metal-rich stars \citep{Zhu2018,Bryan2019,Bryan2024}.
%It is unclear how such a variety of planetary systems arise. 
%Planet formation models need to explain both systems with peas-in-a-pod-type planets and 
%systems with planets showing a larger variety in radii/masses and orbital separations.

Classical planet formation models with planetesimal accretion tend to lead to \textcolor{black}{the} formation of \textcolor{black}{equally spaced} (typically $\sim10$ mutual-Hill-radii apart) 
similar-mass neighbouring planets due to the oligarchic growth stage \citep{Kokubo2002}.  
In this scenario, a diversity in planetary masses in a system is generally attributed to the existence of the snow line.
%\cite{Emsenhuber2021a} studied...did they talk about mass difference in the system???
\cite{Mishra2021} generated a population of planets by using the Generation III Bern Model that is based on planetesimal accretion \citep{Emsenhuber2021a} and 
by applying the Kepler transit survey biases and found that their models reproduce the observed peas-in-a-pod trends in radii, masses, and orbital separations well.
They also suggested that the mass-size similarity is related to each other. 
This latter outcome might contradict recent findings that the similarity in radii does not necessarily indicate the similarity in masses, 
even in the same system \citep[e.g.,][]{Otegi2022,Mamonova2024}.
%Furthermore, although they have not applied any artificial limits on radii and masses, 
%the vast majority of their planets are non-giants with $\lesssim 10\,R_{\oplus}$ and $\lesssim 50\,M_{\oplus}$. 
%Therefore, it is unclear 
\textcolor{black}{They do not discuss} whether their similarity trends extend to giant planets or 
whether the correlations become weaker by including giant planets\textcolor{black}{,} as seen in observations \citep[e.g.,][]{Mamonova2024}.
%
%Furthermore, it is unclear whether their similarity trends extend to giant planets or 
%whether the correlations become weaker by including giant planets as seen in obervations \citep[e.g.,][]{Otegi2022,Mamonova2024}.
%This is because, although they have not applied any artificial limits on radii and masses, 
%the vast majority of their planets are non-giants with $\lesssim 10\,R_{\oplus}$ and $\lesssim 50\,M_{\oplus}$. 
%
%Furthermore, although they have not applied any artificial limits on radii and masses, 
%the vast majority of their planets are non-giants with $\lesssim 10\,R_{\oplus}$ and $\lesssim 50\,M_{\oplus}$. 
%Thus, it is unclear whether their peas-in-a-pod trends extend to giant planets or 
%whether the correlations become weaker by including giant planets as seen in obervations \citep[e.g.,][]{Otegi2022,Mamonova2024}.

In the case of pebble accretion, the protoplanetary cores across the disc are exposed to a similar pebble mass flux since the pebble accretion 
efficiency per protoplanet is generally low \citep[a few \textcolor{black}{percent} or lower\textcolor{black}{;}][]{Ormel2018b}, and thus neighbouring cores tend to have similar masses. 
In this scenario, a diversity in planetary masses can be created when a planet reaching the pebble isolation mass (PIM) halts 
the pebble mass flux for inner protoplanets \citep{Morbidelli2015}.
\cite{Bitsch2019} studied planet formation via pebble accretion in a ``two-alpha'' disc model, which mimics the wind-driven accretion disc 
by assuming two different viscosity $\alpha$ parameters: one for evolving the overall disc structure via mass accretion 
and the other for handling disc physics such as pebble growth or planet migration. 
They showed that the diversity in mass distribution represented by systems of inner \textcolor{black}{super-Earths} and outer gas giants 
can be created for an intermediate pebble flux, 
while only low-mass \textcolor{black}{or} high-mass planets are formed for a lower \textcolor{black}{or} higher pebble mass flux\textcolor{black}{, respectively}.
However, assuming that their adopted range of pebble fluxes is a proxy for the stellar metallicities, 
mixed-mass systems are expected to form only in intermediate metallicity environments. 
Since there appears to be a preference of such systems around metal-rich stars \citep{Zhu2018,Bryan2019,Bryan2024}, 
their model may not explain all the \textcolor{black}{systems with a short-period \textcolor{black}{super-Earth} accompanied by a cold Jupiter}.  
%SE-CJ \textcolor{black}{Minor point 1: short-period super Earth %accompanied by a cold Jupiter} systems. 

The wind-driven accretion discs could lead to a variation in planet formation outcomes depending on the wind efficiency.
As further discussed in Section~\ref{sec:Methods}, the mass accretion rate in the wind-driven accretion disc has a radial dependence 
\textcolor{black}{that} becomes steeper for the \textcolor{black}{`less-efficient'} wind that removes more mass and angular momentum from the disc to have the 
same level of the disc mass accretion rate \citep[e.g,][]{Tabone2022a,Chambers2019}.
Assuming that the pebble mass flux scales with the gas mass flux to a first approximation \citep[e.g.,][]{Birnstiel2012a}, 
such a radial dependence could affect the outcome of planet formation by changing the optimum location for protoplanets to reach the PIM 
or even by creating a larger mass difference between neighbouring protoplanets.
A \textcolor{black}{radially dependent} mass accretion rate also leads to a flatter profile of the surface mass density, which slows migration for non-gap-opening planets \citep{Ogihara2018}.
In this paper, we test the hypothesis that more efficient wind-driven accretion discs tend to lead to similar-mass neighbouring cores \textcolor{black}{that} migrate faster 
than less efficient wind-driven accretion discs and thus result in \textcolor{black}{the} formation of planetary systems similar to the observed peas-in-a-pod systems. 
%close-in, similar-mass multiple-planet systems. 
 
This paper is organised as follows: In Section~\ref{sec:Methods}, we present the wind-driven accretion disc model and the pebble accretion model adopted 
in this work. 
In Section~\ref{sec:Results}, we present the growth tracks of planets in wind-driven discs, 
investigate how the magnetic lever arm $\lambda$ affects planet formation and migration, and compare our simulations with observations. 
%and present an equation to predict the occurrence rate of giant planets based on the residual disc mass and age of an observed disc. 
Finally, we discuss our results further in Section~\ref{sec:Discussion} and summarise our main findings in Section~\ref{sec:Summary}.

% Understanding how protoplanetary discs evolve has long been a key question in the field of planet formation. 
% Protoplanets form within these discs, whose evolution and dissipation fundamentally control the timescale and stages of planet formation. 
% Observations suggest that the lifetimes of protoplanetary discs range from a few million years (Myrs) to several tens of Myrs. 
% The fact that these discs continue to accrete gas and dust has led to the proposal of the 'alpha-disc model,' 
% where disc accretion is driven by turbulence acting as an effective viscosity. 
%
% This paper is organised as follows: In Section~\ref{sec:Methods}, we present the wind-driven disc model and the pebble accretion model adopted 
% in this work. 
% In Section~\ref{sec:Results}, we present the growth tracks of planets in wind-driven discs, 
% investigate how metallicity and $\lambda$ affect planet formation and migration, 
% compare our simulations with observations and present an equation to predict the occurrence rate of giant planets 
% based on the residual disc mass and age of an observed disc. 
% Finally, we discuss our results further in Section~\ref{sec:Discussion} and summarise our main findings in Section~\ref{sec:Summary}.

\section{Methods}
\label{sec:Methods}
In this study, we employed the formalism of \textcolor{black}{\cite{Tabone2022a}} to compute the evolution of discs influenced by both turbulence and disc winds, 
as briefly summarised in section \textcolor{black}{Section}~\ref{subsec:disc-evolution}. 
By adjusting the disc accretion parameters, %$\alpha_{\text{total}}$ and $\alpha_{\text{DW}}$, 
we explore a spectrum of disc evolution scenarios, 
ranging from purely viscously evolving discs to strongly wind-driven accretion discs. 

To study planet formation in such a disc, we consider just one planetary embryo per disc for simplicity.
The planet formation model generally follows the approach by \cite{Matsumura2021}, but we briefly summarise the model here. 
The pebble accretion, gas accretion, and planet migration models are described in \textcolor{black}{Sections}~\ref{subsec: Pebble accretion}, \ref{subsec:gas-envelope-accretion}, and \ref{subsec: Planet migration}, respectively, 
and the initial conditions of simulations are summarised in \textcolor{black}{Section}~\ref{subsec: ICs}.
%For simplicity, in each calculation, we consider only one planetary embryo within the disc. 
%The core subsequently undergoes pebble accretion (\S\ref{subsec: Pebble accretion}), 
%gas accretion (\S\ref{subsec:gas-envelope-accretion}), and planet migration (\S\ref{subsec: Planet migration}). 

\subsection{Disc \textcolor{black}{evolution}}
\label{subsec:disc-evolution}

\subsubsection{Master \textcolor{black}{equation}}
\label{subsubsec:master-equation}
Our study adopted the 1D disc evolution model described by \cite{Tabone2022a}. 
The model is governed by a master equation that encapsulates the interplay between mass accretion driven by turbulent viscosity, parameterised by \(\alpha_{\rm SS}\), 
and angular momentum transport due to magnetohydrodynamic (MHD) disc winds, parameterised by \(\alpha_{\rm DW}\). 
%The master equation also accounts for the magnetic lever arm parameter, \(\lambda\), which is the efficiency of angular momentum extraction by the wind. 
The full expression of the master equation is given as follows for the evolution of the surface mass density $\Sigma$:

% \begin{align}
% \frac{\partial \Sigma}{\partial t} = \frac{3}{r} \frac{\partial}{\partial r} \left[ \frac{1}{r\Omega} \frac{\partial}{\partial r} \left(r^2 \alpha_{SS} \Sigma c_s^2\right) + \right. & \nonumber \\
% \left. \frac{3}{2r} \frac{\partial}{\partial r} \left(\alpha_{DW} \Sigma c_s^2 \Omega\right) - \frac{3 \alpha_{DW} \Sigma c_s^2}{4(\lambda - 1) r^2 \Omega} \right].
% \label{eq:master-equation}
% \end{align}

% \begin{align}
% = \frac{1}{2 \pi r} \frac{\partial}{\partial r} (2 \pi r v_{r,SS} \Sigma) + \frac{1}{2 \pi r} \frac{\partial}{\partial r} (2 \pi r v_{r,DW} \Sigma) - \frac{\partial \Sigma_{DW}}{\partial t}
% \label{eq:master-equation_rearrange}
% \end{align}

\begin{equation}
    % \frac{\partial \Sigma}{\partial t} &=& 
    % \frac{3}{r} \frac{\partial}{\partial r} \left[\frac{1}{r\Omega} \frac{\partial}{\partial r} \left(r^2 \alpha_{SS} \Sigma c_s^2\right) \right] + \nonumber \\
    % &&\frac{3}{2r} \frac{\partial}{\partial r} \left(\alpha_{DW} \Sigma c_s^2 \Omega\right) - \frac{3 \alpha_{DW} \Sigma c_s^2}{4(\lambda - 1) r^2 \Omega} 
    % \label{eq:master-equation} \\
%
    \frac{\partial \Sigma}{\partial t}
    = \frac{1}{2 \pi r} \frac{\partial}{\partial r} (2 \pi r v_{r,{\rm SS}} \Sigma) + \frac{1}{2 \pi r} \frac{\partial}{\partial r} (2 \pi r v_{r,{\rm DW}} \Sigma) - \frac{\partial \Sigma_{\rm DW}}{\partial t},
    \label{eq:master-equation_rearrange}
\end{equation}
where the first and the second terms on the right-hand side represent the surface mass density evolution 
due to the disc mass accretion ($\dot{M}_{\rm acc}=2\pi r v_r\Sigma$) initiated by the turbulence and the disc wind, respectively, 
while the third term shows the mass removal by the wind
\footnote{Each of these terms corresponds to each term on the right-hand side of Equation~10 in \cite{Tabone2022a}, respectively.}.

Here, the radial velocities are defined as follows by using the viscosity $\nu=\alpha c_s^2/\Omega$, 
where \(c_s\) is the local sound speed and \(\Omega\) is the Keplerian angular velocity\textcolor{black}{:}
\begin{eqnarray}
    v_{r,{\rm SS}} &=& \frac{3\nu_{\rm SS}}{r}\left(\frac{1}{2} + \frac{\partial\ln\left(\Sigma\nu_{\rm SS}\right)}{\partial\ln r}\right) \\
    v_{r,{\rm DW}} &=& \frac{3}{2}\frac{\nu_{\rm DW}}{r}\textcolor{black}{.}
\end{eqnarray}
Each viscosity term is defined by using the Shakura-Sunyaev parameter $\alpha_{\rm SS}$ \citep{ShakuraSunyaev1973}
and the coresponding parameter $\alpha_{\rm DW}$ representing the normalised wind torque.

Furthermore, the third term shows the mass removal by the wind as
\begin{equation}
%    \frac{\partial \Sigma_{DW}}{\partial t} = \frac{3 \alpha_{DW} \Sigma c_s^2}{4(\lambda - 1) r^2 \Omega} \, ,
    \frac{\partial \Sigma_{\rm DW}}{\partial t} = \frac{2\pi rv_{r,{\rm DW}} \Sigma}{4\pi(\lambda - 1) r^2} \, ,
\end{equation}
where \(\lambda\) is the magnetic lever arm \textcolor{black}{that} represents the efficiency of angular momentum removal 
by the wind and is defined as the proportion of angular momentum transported 
by the wind along a streamline relative to the amount at its base \citep{BlandfordPayne1982}. 
%The observed values of \(\lambda\) are approximately 5.5 for HH212, according to \citep{Tabone2020}, 
%and approximately 1.7 for L1448-mm, according to \citep{Nazari2024}. 
%Each term plays a significant role in dictating the dynamics and evolution of the disc's structure. 
For all the simulations in this work, we have assumed \(\alpha_{\rm SS}=10^{-4}\) for wind-driven discs.
Assuming both \(\lambda\) and \(\alpha_{\rm DW}\) are constant in time and disc radius, 
Equation~\ref{eq:master-equation_rearrange} can be solved analytically \citep{Tabone2022a}.

\subsubsection{Surface \textcolor{black}{mass density}}
\label{subsubsec: Surface mass density}

%In the presence of a disc wind, we can characterise the surface density's power-law profile as
%Following Appendix C of \cite{Tabone2022a}, we characterise the surface density's power-law profile as, 
%\begin{equation}
%\Sigma(r) \propto r^{-\gamma+\xi} \, ,
%\label{eq:sigma power-law}
%\end{equation}
Following Appendix C of \cite{Tabone2022a}, the evolution of the surface mass density is described by the following equation:
\begin{equation}
\Sigma(r, t) = \Sigma_c(t) \left(\frac{r}{r_c(t)}\right)^{-\gamma+\xi} \exp\left[-\left(\frac{r}{r_c(t)}\right)^{2-\gamma}\right],
\label{eq:sigma_r_t}
\end{equation}
where \(\gamma\) is the exponent that describes the radial dependency in a disc that is entirely viscously evolving. 
%This exponent is set to \(\frac{15}{14}\), 
Throughout this paper, we have adopted $\gamma=15/14$, a value that aligns with the exponent of the surface mass density in the irradiation-dominated region of the disc model by \cite{Ida2016}. 
%\textcolor{black}{Minor point 2: This value arises from $q_2 = \frac{3}{7}$, which is the exponent in the temperature profile $T \propto r^{-q_2}$ in the irradiation regime \citep{Ida2016}}.

The parameter \(\xi\equiv\frac{d\ln\dot{M}_{\rm acc}}{d\ln r}\), known as the wind's mass ejection index, can be expressed as 
\begin{equation}
\xi = \frac{(\psi + 1)}{4} \left[ \sqrt{1 + \frac{4\psi}{(\lambda - 1)(\psi + 1)^2}} - 1 \right] \, .
\label{eq:xi}
\end{equation}
This index quantifies the local mass-loss rate relative to the local accretion rate, as defined by \cite{FerreiraPelletier1995}, 
while \(\psi\) represents the relative strength between the radial and the vertical torque \citep{Tabone2022a} as
\begin{equation}
\psi = \frac{\alpha_{\rm DW}}{\alpha_{\rm SS}} \, .
\label{eq:psi}
\end{equation}
%
% The evolution of the surface mass density is described by the following equation:
% %
% \begin{equation}
% \Sigma(r, t) = \Sigma_c(t) \left(\frac{r}{r_c(t)}\right)^{-\gamma+\xi} \exp\left[-\left(\frac{r}{r_c(t)}\right)^{2-\gamma}\right],
% \label{eq:sigma_r_t}
% \end{equation}
% %
% where $\Sigma_c(t)$ represents the surface density at the disc's outer edge, evolving over time. 
% The parameter $r_c(t)$ denotes the initial characteristic radius of the disc, impacting the radial extent and the accretion rate. 

Furthermore, \textcolor{black}{$r_c(t)$ represents the exponential cut-off radius as seen in Equation~\ref{eq:sigma_r_t}, which can also be interpreted as the disc's outer edge. The evolution of $r_c(t)$} is governed by the equation:
\begin{equation}
r_c(t) = r_c(0) \left(1 + \frac{t}{(1 + \psi)t_{\text{acc},0}}\right)^{\frac{1}{2-\gamma}},
\label{eq:r_c_t}
\end{equation}
where $r_c(0)$ is the initial characteristic radius while $t_{\text{acc},0}$ is the initial accretion timescale that is defined as, 
%\textcolor{black}{Minor point 3: which represents the exponential cut-off radius. In wind-driven discs, where viscous spreading is negligible, $r_c$ can also be considered as the disc's outer edge.} 
%
\begin{equation}
t_{\text{acc},0} \equiv \frac{r_c(0)}{3\textcolor{black}{(2-\gamma)^2} \hat{h}_{g,c} c_{s,c} \alpha_{\rm total}} \, .
\label{t_acc}
\end{equation}
%
%\(\tilde{\alpha} = \alpha_{DW} + \alpha_{SS}\)
Here, \(\alpha_{\rm total} = \alpha_{\rm DW} + \alpha_{\rm SS}\) is the \(\alpha\)-parameter that quantifies the total torque exerted by the MHD disc wind and turbulence, 
while \(c_{s}\) denotes the speed of sound with the subscript \(c\) indicating that the parameter is evaluated at \textcolor{black}{$r_c$}.
Furthermore, \(\hat{h}_g = \frac{h_g}{r}\) represents the disc's aspect ratio in the irradiation regime \citep{Ida2016}\textcolor{black}{:}
%, with the formulation derived from the irradiation regime 
%as presented in \citep{Ida2016}. 
%
\begin{equation}
\hat{h}_{g} \approx 0.024L_{*0}^{1/7}M_{*0}^{-4/7} \left( \frac{r}{1 \, \text{au}} \right)^{2/7}\textcolor{black}{.}
\end{equation}
\textcolor{black}{Here, the exponent $\beta=2/7$ arises from $q = 3/7$, which is the exponent in the temperature profile $T \propto r^{-q}$ in the irradiation regime \citep{Ida2016}. For a purely \textcolor{black}{viscously evolving} steady disc, these parameters are related as $\gamma=-q+3/2=2\beta+1/2$.}

Finally, the temporal evolution of the surface density at $r_c$ is expressed as
%
%\begin{equation}
%\Sigma_c(t) = \Sigma_c(0) \left(1 + \frac{t}{(1 + \psi)t_{\text{acc},0}}\right)^{-\frac{(5+2\xi+\psi)}{2(2-\gamma)}}
%\end{equation}
%
%\begin{equation}
%    \Sigma_c(t) = \frac{M_D(t)}{2\pi r_c(t)^2} = \Sigma_c(0) \left(1 + \frac{t}{(1 + \psi)t_{\text{acc},0}}\right)^{-\frac{(5+2\xi+\psi)}{2(2-\gamma)}} \, .
%\end{equation}
%
%\begin{eqnarray}
%    \Sigma_c(t) &=& \Sigma_c(0) \left(1 + \frac{t}{(1 + \psi)t_{\text{acc},0}}\right)^{-\frac{(5+2\xi+\psi)}{2(2-\gamma)}} \\
%    \Sigma_c(0) &=& \frac{M_D(0)}{2\pi r_c(0)^2} \, ,
%\end{eqnarray} 
%
\begin{equation}
    \Sigma_c(t) = \Sigma_c(0) \left(1 + \frac{t}{(1 + \psi)t_{\text{acc},0}}\right)^{-\frac{(5+2\xi+\psi)}{2(2-\gamma)}} \, ,
\end{equation} 
where $\Sigma_c(0)$ is related to the initial disc mass $M_D(0)$ as follows:
\begin{equation}
    \Sigma_c(0) = \frac{M_D(0)}{2\pi r_c(0)^2}\textcolor{black}{.}
\end{equation}

\begin{figure}[ht]
\centering
\includegraphics[width=\linewidth]{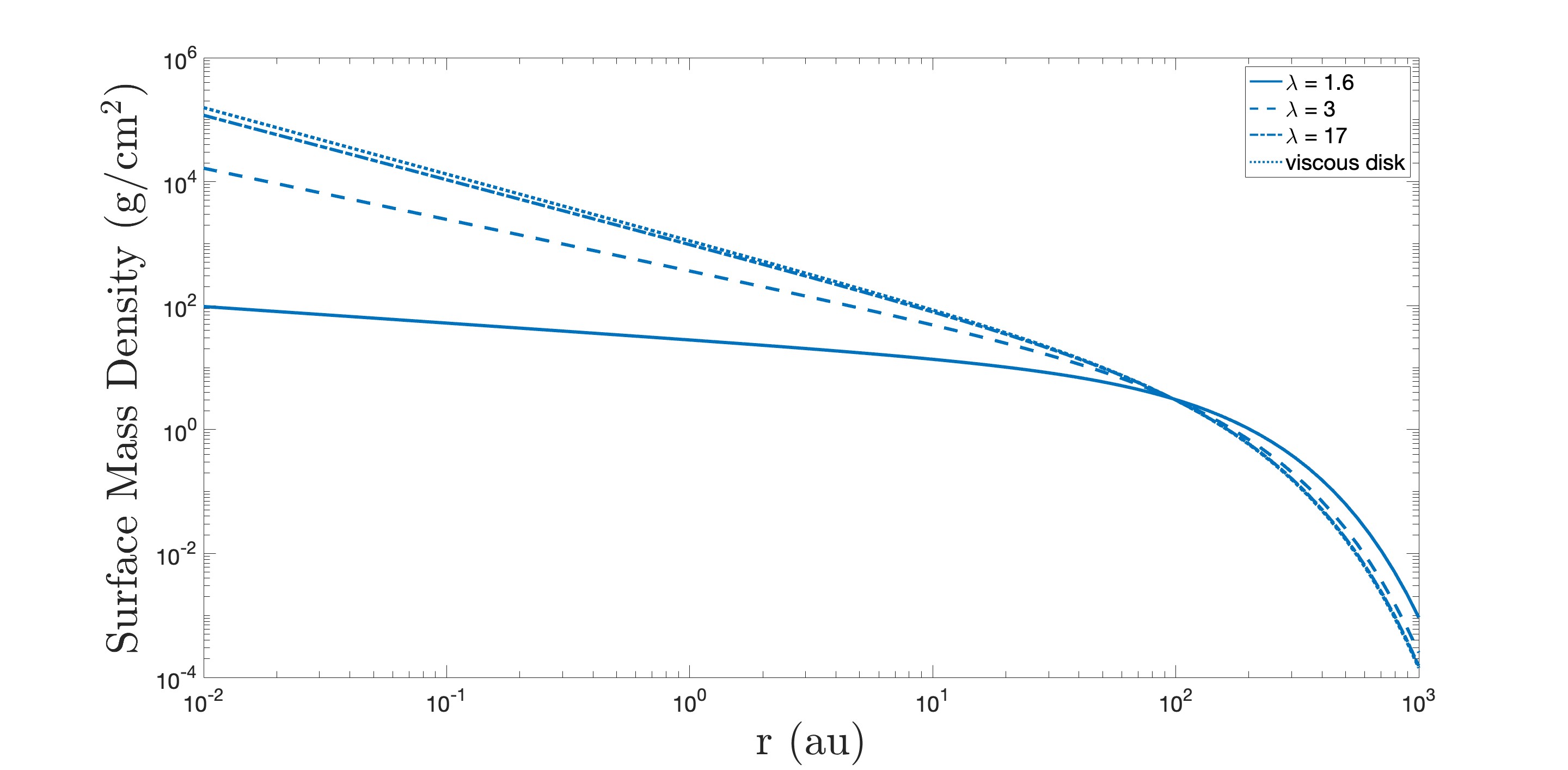}
\caption{\textcolor{black}{Initial} surface mass densities of Disc 5 for 
WD discs with $\lambda=1.6$, $3$, and $17$ (solid, dashed, and dash-dotted, respectively) compared with 
that for the VE disc (dotted). 
More inefficient WD discs lead to flatter surface mass densities due to the stronger wind mass loss.
%\textcolor{black}{We note} that 
\textcolor{black}{All} the discs have initially the same mass.}
\label{fig:sigma0disc5}
\end{figure}

The surface mass density profile is controlled by $\xi$ via both \(\lambda\) and \(\psi\). 
In Figure~\ref{fig:sigma0disc5}, we show how different $\lambda$ and $\psi$ lead to 
different surface mass density profiles for the same disc mass of 0.06 \(M_{\odot}\).
Here, the dotted line represents the purely viscously evolving (VE) disc, while the dash-dotted, dashed, and solid lines correspond to 
the wind-driven (WD) accretion discs with \(\lambda = 17\), \(\lambda = 3\), and \(\lambda = 1.6\), respectively.

For VE discs, $\psi=0$ because $\alpha_{\rm DW}=0$, while for WD discs, $\psi=7.4\times10^{-3}/10^{-4}=74$ is assumed here 
%\textcolor{black}{$\psi=7.4\times10^{-3}/10^{-4}=74$?} is assumed here 
so that the disc accretion is dominated by the winds rather than by the turbulence.
The choices of the $\lambda$ values are motivated by Figure~3 of \cite{Tabone2022a}, 
where \(\lambda = 17\) corresponds to \(\xi \approx 0\) in the wind-dominated case, 
\(\lambda = 1.6\) has the highest mass ejection index of \(\xi \approx 0.8\), 
and \(\lambda = 3\) represents the intermediate case of \(\xi \approx 0.2\). 
The observed values of \(\lambda\) are $\sim5.5$ for HH212 \citep{Tabone2020}, 
and $\sim1.7$ for L1448-mm \citep{Nazari2024}, so $\lambda=17$ may be unrealistic.
However, we have included this case to compare the planet formation outcome with VE discs.

% Figure~\ref{fig:sigma0disc5} shows the initial surface mass density profiles for the disc mass of 0.06 \(M_{\odot}\) for four different disc models. 
% The dotted line represents the viscously evolving disc, while the dash-dotted, dashed, and solid lines correspond to 
% \(\lambda = 17\), \(\lambda = 3\), and \(\lambda = 1.6\), wind-driven discs respectively. 
% In Figure~3 of \cite{Tabone2022a}, \(\lambda = 17\) corresponds to \(\xi \approx 0\) in the wind-dominated case, 
% \(\lambda = 1.6\) has the highest mass ejection index, and \(\lambda = 3\) represents \(\xi \approx 0.2\). 
% Therefore, these three different \(\lambda\) values are selected in our simulations to represent both mild and intense wind-ejection discs.
As can be seen in Figure~\ref{fig:sigma0disc5}, the smaller the value of \(\lambda\) is, the flatter the slope of \(\Sigma\) becomes.
\cite{Ogihara2018} studied the orbital evolution of type I migrators in the WD accretion discs and found that the type I migration is 
significantly suppressed when the surface mass density becomes flatter.  
Thus, lower $\lambda$ discs are expected to lead to less efficient type I migration. 

\textcolor{black}{Here, we note that the WD accretion model by \cite{Tabone2022a} does not lead to the formation of an inner disc cavity as shown by \cite{Suzuki2016}, because their model assumes uniform $\alpha$ and $\lambda$ parameters rather than the spatial and temporal evolution of these as in \cite{Suzuki2016}.  As a result, the adopted disc model does not lead to the positive slope in the surface mass density that can reverse planet migration.}
%\textcolor{black}{\cite{Sukuzi2016} also studied the evolution of surface mass density in WDs using a sophisticated energetic approach, predicting an inverse slope of $\Sigma$ and the formation of cavities. In contrast, this work adopts the WD model by \cite{Tabone2022a} with simplified assumptions, such as a constant $\lambda$ parameter both spatially and temporally, resulting in a negative slope in $\Sigma$. Consequently, investigating features such as planet formation in WDs where the slope of $\Sigma$ is inverted lies beyond the scope of this paper. Instead, we focus on scenarios where the slope of $\Sigma$ remains negative.}

\subsubsection{Disc and \textcolor{black}{stellar accretion}}
\label{subsubsec: Disc and stellar accretion}

In \cite{Tabone2022a}, the local mass accretion rate varies across the disc due to the wind mass loss as 
\begin{equation}
    \dot{M}_{\rm acc}\left(r,\,t\right) = \dot{M}_{*}\left(t\right)\left(\frac{r}{r_{{\rm in}}}\right)^{\xi} 
     = \dot{M}_{D}\left(t\right)\left(\frac{r}{r_{c}\left(t\right)}\right)^{\xi}  \, ,
    \label{eq:dotMacc}
\end{equation}
where $\dot{M}_{*}$ and $\dot{M}_{D}$ are the stellar accretion rate and the disc accretion rate, respectively, and 
are defined at the inner wind launching radius $r_{\rm in}$ and the outer wind launching radius $r_{c}$, respectively, so that 
\begin{eqnarray}
    \dot{M}_{{\rm acc}}\left(r_{{\rm in}},\,t\right)&=&\dot{M}_{*}\left(t\right) \label{eq:dotMs}\\
    \dot{M}_{{\rm acc}}\left(r_{c},\,t\right)&=&\dot{M}_{D}\left(t\right) \label{eq:dotMD}\, .
\end{eqnarray}
%
%In the model developed by \cite{Tabone2022a}, the disc mass is given as a function of time as, 
%as a function of time is given as:
% The disc mass is given as a function of time as,
% %
% \begin{equation}
% M_D(t) = M_0 \left( 1 + \frac{t}{(1 + \psi)t_{\text{acc},0}} \right)^{-\frac{(1+2\xi+\psi)}{2(2-\gamma)}} \, ,
% \label{eq:disc_mass}
% \end{equation}
% %
% where \( M_0 \) is the initial disc mass. %the rate of mass loss in the disc can be described as the time derivative of Equation \ref{eq:disc_mass}, as follows:
% Using this, the disc mass accretion rate is obtained as follows.
%
%\begin{equation}
%\dot{M_D} = -\frac{M_D \cdot \left(\xi + \frac{\psi}{2} + \frac{1}{2}\right)}{t_{acc,0} \cdot \left(\frac{t}{t_{acc,0} \cdot (\psi + 1)} + 1\right)^{\xi + \psi/2 + 3/2} \cdot (\psi + 1)}
%\label{eq:dot(M_D)}
%\end{equation}
%
The disc mass accretion rate is obtained as follows from Equation~(C8) of \cite{Tabone2022a}:
\begin{eqnarray}
    \dot{M}_{D}\left(t\right) &=& \dot{M}_{D}\left(0\right)\left(1+\frac{t}{\left(1+\psi\right)t_{{\rm acc},0}}\right)^{-\left(5-2\gamma+2\xi+\psi\right)/\left(2\left(2-\gamma\right)\right)} \\ 
    \dot{M}_{D}\left(0\right) &=& \textcolor{black}{-}\frac{\left(1+2\xi+\psi\right)}{\left(1+\psi\right)}\frac{M_D(0)}{2\left(2-\gamma\right)t_{{\rm acc},0}} \, .
\end{eqnarray}
%
%where \( M_0 \) is the initial disc mass.
%
We can use this expression of $\dot{M}_{D}$ in Equation~\ref{eq:dotMacc} to calculate the local mass accretion rate $\dot{M}_{{\rm acc}}$.

Alternatively, we can find the experssion of $\dot{M}_*$ to calculate $\dot{M}_{{\rm acc}}$.
The local mass accretion rate $\dot{M}_{{\rm acc}}$ is related to the cumulative wind mass-loss rate $\dot{M}_{w}$ as 
\begin{equation}
    \dot{M}_{{\rm acc}}\left(r,\,t\right) = \dot{M}_{w}\left(r,\,t\right) + \dot{M}_{*}\left(t\right) \, .
    \label{eq:Macc}
\end{equation}
\cite{Tabone2022a} defines the mass ejection-to-accretion ratio as 
\begin{equation}
f_M(t) \equiv \frac{\dot{M}_{w}\left(r_c,\,t\right)}{\dot{M}_{*}} = \left(\frac{r_c(t)}{r_{\text{in}}}\right)^\xi - 1 \, .
\label{eq:f_M_t}
\end{equation}
%
%and writes the stellar mass accretion rate as the disc mass flux \(\dot{M}_D\) reduced by the fraction mass that is effectively accreted onto the growing star:
Using this definition and evaluating Equation~\ref{eq:Macc} at $r=r_{c}$, the stellar mass accretion rate is written 
as the disc mass flux \(\dot{M}_D\) reduced by the fraction mass that is effectively accreted onto the growing star:
\begin{equation}
\dot{M}_*(t) = \frac{1}{1 + f_M(t)} \dot{M}_D(t).
\label{eq:stellar_accretion_rate}
\end{equation}
%
%With the main sources being disc wind mass loss and stellar accretion.  
% The exact contribution is also quantified through the mass ejection-to-accretion ratio :
%
% \begin{equation}
% f_M(t) = \left(\frac{r_c(t)}{r_{\text{in}}}\right)^\xi - 1
% \label{eq:f_M_t}
% \end{equation}
%
% Where \( r_{\text{in}} \) is the radial location at which the disc wind is launched, 
% we assume that this value is fixed at 1~\text{AU} and does not depend on time.
%
% In the presence of the disc wind, the stellar accretion rate \(\dot{M}_*\) is the disc mass flux \(\dot{M}_D\) 
% reduced by the fraction mass that is effectively accreted onto the growing star:
%
% \begin{equation}
% \dot{M}_*(t) = \frac{1}{1 + f_M(t)} \dot{M}_D(t).
% \label{eq:stellar_accretion_rate}
% \end{equation}
%
% Due to the local mass removal by the wind, the local accretion rate $\dot{M}_{\text{acc}}(r)$, is described as:
%
% \begin{equation}
% \dot{M}_{\text{acc}}(r) = \dot{M}_* \left( \frac{r}{r_{\text{in}}} \right)^{\xi}.
% \label{eq:local_accretion_rate}
% \end{equation}

\begin{figure}[ht]
    \centering
    \includegraphics[width=\linewidth]{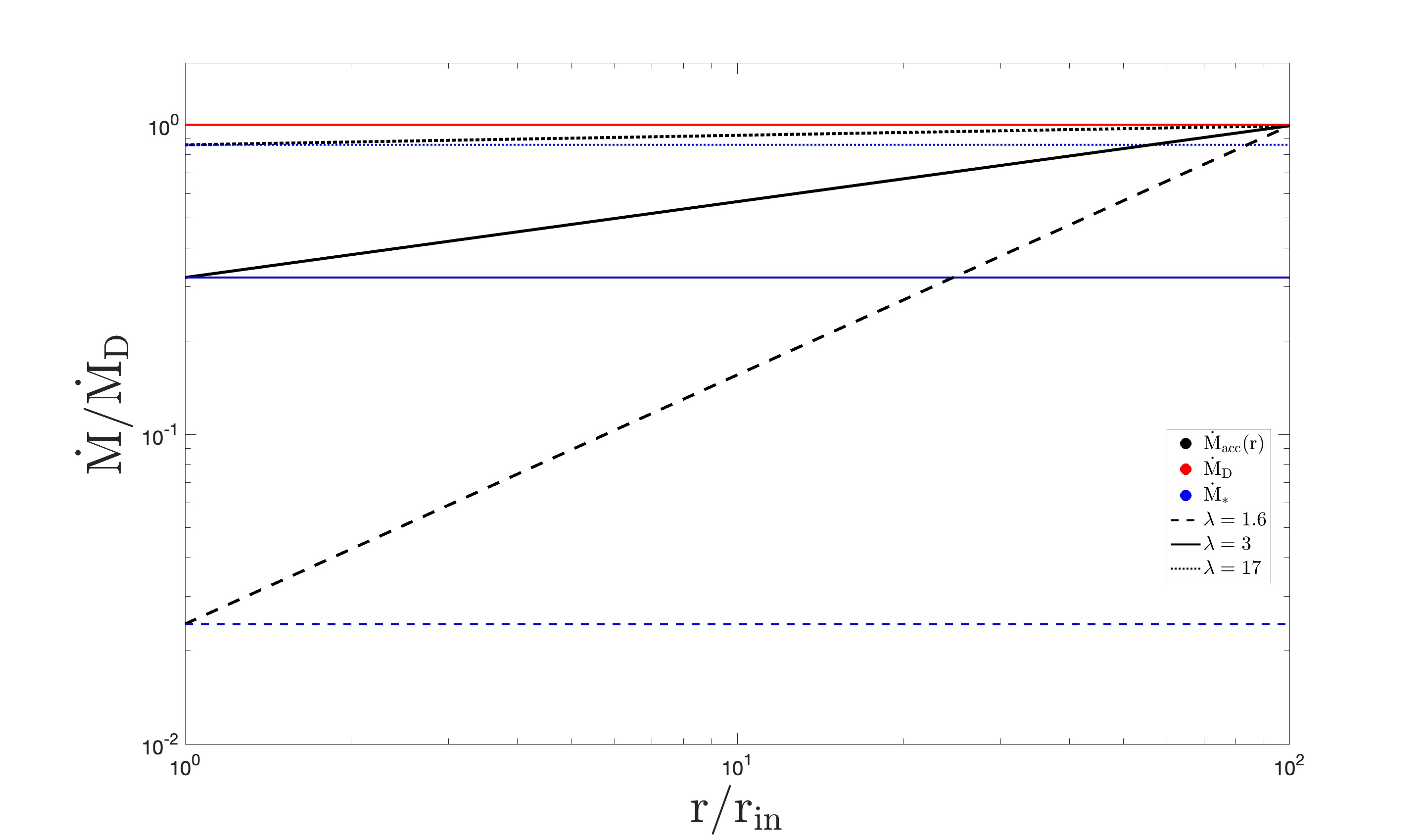}
    %\caption{The radial profile of Disc local accretion rates, disc mass loss rates and stellar accretion rates for disc 5.}
    \caption{\textcolor{black}{Radial} profiles of the local disc accretion rates $\dot{M}_{\rm acc}$ normalised to the disc accretion rate \(\dot{M}_D\) are shown 
    for Disc 5 with $\lambda=1.6$ (black dashed line), $\lambda=3.0$ (black solid line), and $\lambda=17$ (black dotted line).  
    Also plotted are the normalised stellar mass accretion rate $\dot{M}_*$ (\textcolor{black}{blue lines}) and the normalised disc mass loss rates $\dot{M}_D$ (\textcolor{black}{red line}). 
    For the same disc model, $\dot{M}_D$ is the same for all values of $\lambda$. 
    Due to the substantial wind mass loss, the smaller $\lambda$ leads to the steeper radial dependence of $\dot{M}_{\rm acc}$.
    }
    \label{fig:local_acc_rate}
\end{figure}
%
%Figure \ref{fig:local_acc_rate} illustrates the radial dependence of \(\dot{M}_{\text{acc}}\), \(\dot{M}_D\), and \(\dot{M}_*\) 
%normalised to \(\dot{M}_*\) with \(\lambda = 1.6, 3,\) and \(17\). 
Figure~\ref{fig:local_acc_rate} illustrates the radial dependence of \(\dot{M}_{\text{acc}}\) normalised by \(\dot{M}_D\) 
for \(\lambda = 1.6, 3,\) and \(17\) (black dashed, solid, and dotted lines, respectively). 
Also plotted are normalised \(\dot{M}_D\) (red line) and \(\dot{M}_*\) (blue lines).
As expected from Equations~\ref{eq:dotMs} and \ref{eq:dotMD}, \(\dot{M}_{\text{acc}}\) agrees with \(\dot{M}_*\) and \(\dot{M}_D\) at \(r_{\text{in}}\) and \(r_c\), respectively.
%A smaller \(\lambda\) corresponds to a more significant wind mass-loss, thereby inducing a steeper radial dependence of the local accretion rate.
%As already shown in Figure~\ref{fig:sigma0disc5}, such a mass loss leads to a flatter surface mass density.
%
%Between \(r_{\text{in}}\) and \(r_{\text{out}}\), \(\dot{M}_{\text{acc}} \propto r^\xi\) and is equal to \(\dot{M}_D\) at \(r = r_{\text{out}}\). 
%A smaller \(\lambda\) significantly reduces disc mass, thereby inducing a stronger local accretion rate, 
%which can have a substantial impact on planet formation and migration.

The disc accretion rate of the VE disc is not explicitly shown, but corresponds to the red horizontal line in Figure~\ref{fig:local_acc_rate}, 
because the local disc accretion rate there is independent of the radius\textcolor{black}{,} and thus \(\dot{M}_{\text{acc}}=\dot{M}_*=\dot{M}_D\).
For the WD discs, the local disc accretion rate has a radial dependence and the slope becomes steeper for a smaller $\lambda$ 
due to a more substantial mass loss. 
Furthermore, since the value of $\dot{M}_D$ is the same for all the cases, the stellar mass accretion rate decreases 
for the lower $\lambda$.

\begin{figure}[ht]
    \centering
    \includegraphics[width=\linewidth]{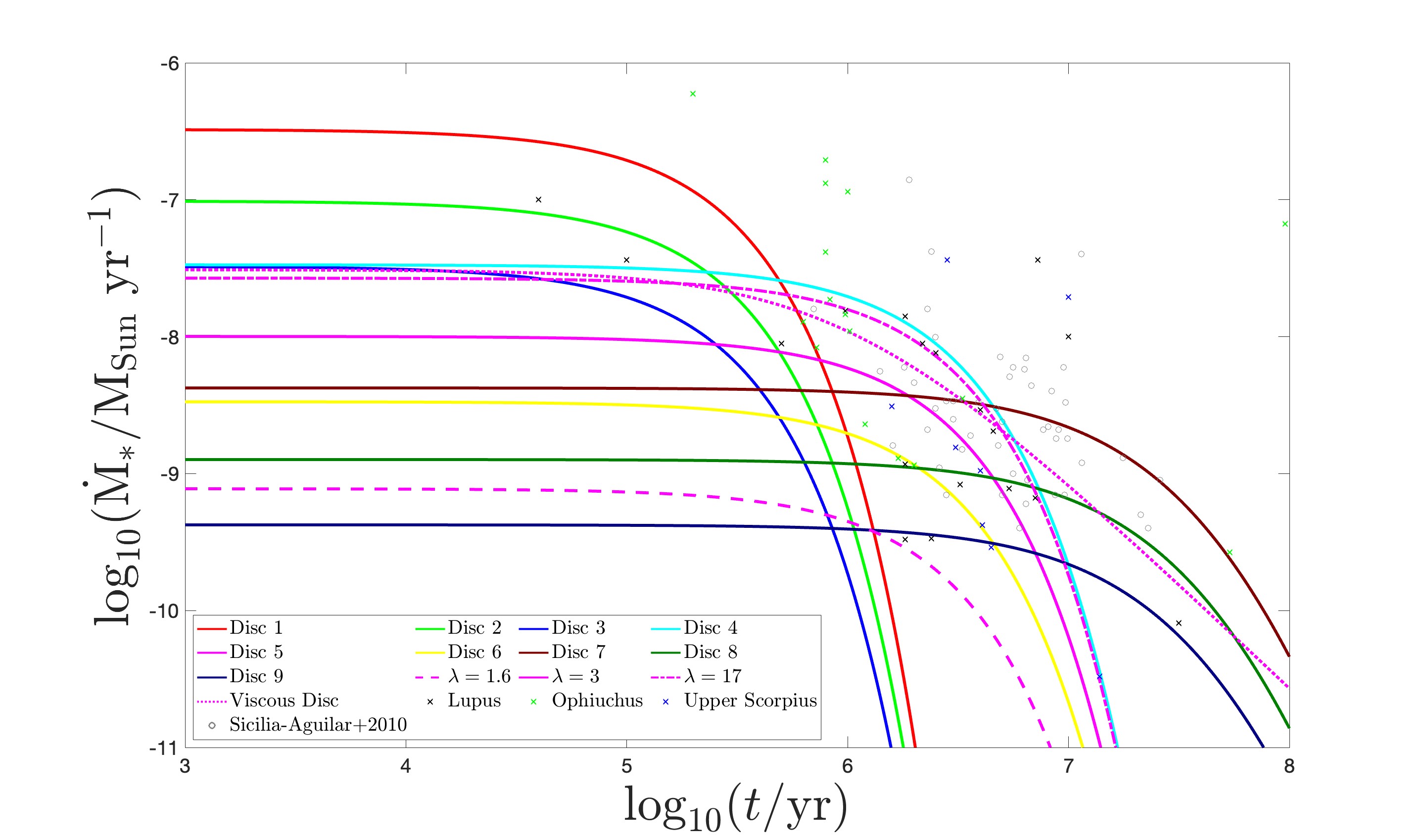}
    \caption{\textcolor{black}{Temporal} profile of stellar accretion rates for the nine discs studied in this work (Table \ref{tab:disc_parameters}) when \(\lambda = 3\). 
    Additionally, cases with \(\lambda = 1.6\), \(\lambda = 17\), and purely \textcolor{black}{viscously evolving} scenarios for disc 5 are also shown.
    Various symbols correspond to observed values: black cross markers:\cite{Batalha1993};\cite{Galli2015};\cite{Nuernberger1997};\cite{Andrews2018};\cite{Alcala2014}; 
    Green cross markers:\cite{Andrews2018};\cite{Erickson2011};\cite{Andrews2010}; 
    Blue cross markers:\cite{Herczeg2014};\cite{Garufi2020};Sicilia- Aguilar+2020:\cite{SiciliaAguilar2010}}
    \label{fig:stellar_acc_rate}
\end{figure}

\begin{figure}[ht]
\centering
\includegraphics[width=\linewidth]{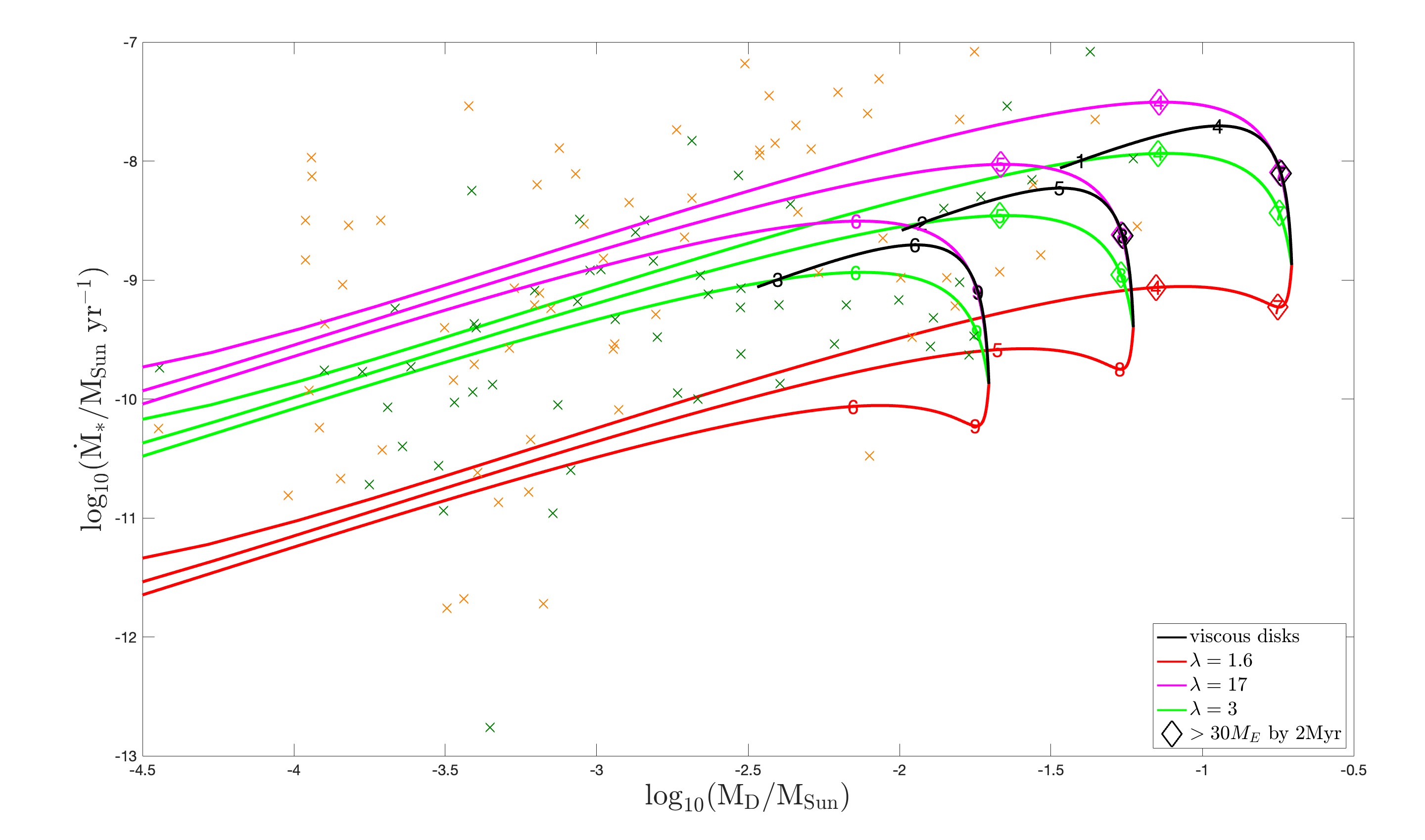}
\caption{\textcolor{black}{Relation of $M_D-\dot{M}_*$ for the WD and VE discs} at 2 Myrs for Discs 1-9 compared 
with the corresponding distributions in Lupus (green) and Chamaeleon I (orange) star forming regions. 
The observational data are taken from \cite{Manara2019}. 
The disc numbers with diamonds imply that the discs produced giant planets ($\geq30 \, M_{\oplus}$) within 2 Myrs.}
\label{fig:disc_isochrones}
\end{figure}

Next, we compare the evolution of the adopted disc models with 
the observed stellar mass accretion rates (Figure~\ref{fig:stellar_acc_rate}) 
and the observed distribution of the disc mass and the stellar mass accretion rate (Figure~\ref{fig:disc_isochrones}).  
Instead of randomly choosing disc parameters, we have selected nine disc models that have a combination of 
the initial disc masses of $0.2M_{\odot}$, $0.06M_{\odot}$, or $0.02M_{\odot}$ and 
the total alpha parameters of $\alpha_{\rm total}=7.4\times10^{-2}$, $7.4\times10^{-3}$, or $7.4\times10^{-4}$ (see Table~\ref{tab:disc_parameters}). 
One of the reasons for this choice is that these are the same set of parameters used by \cite{Matsumura2021}\textcolor{black}{, so it makes} the comparison easier. 
Another reason is that choosing a particular set of parameters rather than the random ones makes it easier to highlight the effects caused by 
a single parameter such as $\lambda$.  
%We will show the results of planet formation derived from a random set of disc parameters in Section~\ref{subsec: Omega}.

Figure~\ref{fig:stellar_acc_rate} compares the stellar accretion rates $\dot{M}_*$ as a function of time 
for Discs 1-9 that are calculated from Equation~\ref{eq:stellar_accretion_rate} 
with the observed stellar accretion rates for different ages that are shown in various symbols (see the caption for references).
%
%Figure~\ref{fig:stellar_acc_rate} shows the stellar accretion rate as a function of time, 
%calculated from equation \ref{eq:stellar_accretion_rate} for discs 1-9 (see Table \ref{tab:disc_parameters}). 
%All the solid lines represent wind-driven discs with \(\alpha_{\text{SS}} = 10^{-4}\) and \(\lambda = 3\). 
For all the disc models shown in solid lines, we have chosen \(\alpha_{\text{SS}} = 10^{-4}\) and \(\lambda = 3\).
For Discs 1-3, for example, $\alpha_{\rm total}=7.4\times10^{-2}$ for all, and the initial disc masses are $0.2M_{\odot}$, $0.06M_{\odot}$, 
and $0.02M_{\odot}$, respectively.
For these discs, Disc 1 has the highest mass accretion rate while Disc 3 has the lowest, 
but the overall disc evolution timescales are similar and discs are gone within $\sim2\,$Myr.
For Discs 4-6 and Discs 7-9, $\alpha_{\rm total}=7.4\times10^{-3}$ and $7.4\times10^{-4}$, respectively.  
The trends for these groups of discs are similar to Discs 1-3, but the disc lifetimes become longer due to the smaller alpha values.
Overall, although each of these disc models has a comparable $\dot{M}_*$ to some of the observed systems, 
none of them produce the highest observed $\dot{M}_*$ for different ages. 

For Disc 5 (magenta line), we have also plotted the purely viscously evolving disc (dotted line) 
and the cases for \(\lambda = 1.6\) (dashed line) and \(17\) (dash-dotted line) for comparison. 
As expected from Figure~\ref{fig:local_acc_rate}, the lower $\lambda$ disc has the lower stellar accretion rate due to a more substantial mass loss.  
%The purple asterisks represent the observed stellar accretion rates from \citep{SiciliaAguilar2010}'s Table 2. 
%Vairous symbols correspond to the observed stellar accretion rates from ***
%All the simulations assume the solar-mass star as well. 
%We note that we modelled solar mass stars in all the simulations, while the observations include GKM stars. 
Since we don't take account of the photoevaporation effect for the VE disc, it has a very long disc lifetime. 
%
%Wind-driven discs exhibit a notably faster depletion rate compared to viscous discs, typically characterised 
%by a rapid decline commencing within a few Myrs. 
% Although wind-driven discs with small \(\lambda\) values have relatively low stellar accretion rates, 
% the local accretion rates in the outer discs could be higher by a factor of a few to several tens. 
% Therefore, active planet formation may be possible in wind-driven discs around stars with low stellar accretion rates. 
%
%%% NOT RELEVANT HERE
% The terms \(L_{\ast0} = \frac{L_{\ast}}{L_{\odot}}\) and \(M_{\ast0} = \frac{M_{\ast}}{M_{\odot}}\) 
% represent the stellar luminosity and mass, respectively, normalised to solar values. 
% The parameter \(\alpha_3 = \frac{\tilde{\alpha}}{10^{-3}}\) is the total disc accretion efficiency \(\alpha\) 
% scaled to a typical value. Additionally, \(\dot{M}_{\ast8} = \frac{\dot{M}_{\ast}}{10^{-8} M_{\odot} \text{yr}^{-1}}\) 
% denotes the stellar mass accretion rate normalised to a standard rate.

Figure~\ref{fig:disc_isochrones} compares the $M_D-\dot{M}_*$ relation of WD and VE discs at 2 Myrs for Discs 1-9 
with the observed distributions of $M_D-\dot{M}_*$ in Lupus and Chamaeleon I star forming regions 
\citep[shown in green and orange crosses, respectively, taken from][]{Manara2019}.   
%for initial masses \(M_0 = 0.02\), \(0.06\), and \(0.2 \, M_{\odot}\). 
%traces the path in the \(M - \dot{M_*}\) plane for a collection of discs, each starting with the same mass \(M\), 
%but varying in \(\tilde{\alpha}\). 
An isochrone is drawn for the evolution of the same disc mass (and the same $\lambda$ for WD discs) but with different $\alpha_{\rm total}$, 
which changes from \(10^{-4}\) to \(10^{-1}\) \textcolor{black}{to encompass the range of $\alpha_{\rm total}$ for our disc models.}  
We have also marked the cases with $\alpha_{\rm total}$ corresponding to Discs 1-9, with diamond symbols highlighting those capable of producing 
at least one giant planet with \(\geq30 \, M_{\oplus}\) within 2 Myrs. 
%while \(\alpha_{\text{SS}}\) is consistent at \(10^{-4}\) across all WD discs. 
%Numerical labels on the isochrones indicate the position of each disc at \(t = 2\) Myrs, with diamond symbols highlighting those capable of producing 
%at least one planet greater than \(30 \, M_E\) by 2 Myrs. 
Discs 1-3 for WD discs are absent because these discs deplete by 2 Myrs (\(M_D < 10^{-4.5} \, M_{\odot}\)), as seen in Figure~\ref{fig:stellar_acc_rate}. 
%The orange and dark green markers denote observations from \citep{Manara2019}. 
WD discs, with their rapid disc drainage capability, cover a broader range of low-disc-mass and low-stellar-accretion rate parameters 
compared to VE discs. 
The figure illustrates that the evolutionary paths of WD discs align reasonably well with the observational data with comparable ages.
However, as in Figure~\ref{fig:stellar_acc_rate}, none of the disc models reproduce highest mass accretion rates over a range of disc masses.
%We will further discuss this point in Section~\ref{subsec: Omega}.

% Due to the presence of the disc wind, the local accretion rate is no longer constant in \( r \); therefore, the steady disc condition 
% \begin{equation}
% \dot{M}_{\text{acc}} = 3 \pi \Sigma_g \nu 
% \label{eq:steady_disc}
% \end{equation}
% must be justified. However, it can be shown that in a wind-driven disc, 
% \begin{equation}
% \dot{M}_{\text{acc}}(r, t) = 3 \pi A \Sigma(r, t) \nu(r, t)
% \label{eq:wind_disc}
% \end{equation}
% %
% As the disc transitions from being purely viscously evolving to entirely wind-driven, \(\xi\) varies from 0 to 1. 
% Therefore, with the choice of \(\lambda = 1.6, 3, \text{ and } 17\), the additional factor
% %
% \begin{equation}
% A = \left( \frac{1 + 2\xi + \psi}{1 + \psi} \right) \Gamma \left( \frac{\xi + 2 - \gamma}{2 - \gamma} \right)
% \label{eq:wind_disc_factor}
% \end{equation}
% %
% ranges from 0.9 to 1.2. An equation similar to \ref{eq:steady_disc} still holds for wind-driven discs.

\subsection{Planet \textcolor{black}{formation via pebble accretion}}
\label{subsec: Pebble accretion}

%Pebble accretion is a novel mechanism in the formation of planetary systems. 
%It involves the gradual accumulation of small, solid particles, or 'pebbles', which significantly contribute to the growth of protoplanetary cores. 
%This process is influenced by a complex interplay of various factors, such as the properties of the disc, the central star, and the evolving planetary bodies.

\subsubsection{Pebble accretion}
\label{subsubsec: Pebble mass flux}
Pebble accretion is a novel mechanism in the formation of planetary systems, which 
involves the gradual accumulation of \textcolor{black}{centimetre to metre-sized particles
} called `pebbles'. 
We have estimated the pebble mass flux, \( \dot{M}_F \), by using the `pebble formation front' model proposed by \cite{LambrechtsJohansen2014}. 
As dust particles aggregate and increase in size to form pebbles, they become highly susceptible to the influence of the headwind. 
This results in a pronounced inward migration of pebbles, supplying a substantial amount of solid mass to the inner regions of the disc. 
Generally, as the timescale for dust growth is longer at greater radii, the origin of the pebble mass flux, 
often referred to as the pebble formation front, progressively shifts \textcolor{black}{outwards} over time. 
We have computed the mass flux of pebbles swept by the pebble formation front by following \cite{Matsumura2021}:
%utilised the formulae in \citep{Matsumura2021} as follows:
%
% \begin{multline}
% \dot{M}_F = 2.066 \times 10^{-2} 
% \left( \frac{\ln \frac{R_{\text{peb}}}{R_0}}{\ln 10^4} \right)^{-1}
% \left( \frac{T_2}{150 \text{ K}} \right)^{-1}
% L_{\ast0}^{-\frac{2}{7}} 
% M_{\ast0}^{\frac{8}{7}} 
% \alpha_3^{-1} 
% \dot{M}_{\text{acc8}} \\
% \left( \frac{\Sigma_{\text{pg,0}}}{0.01} \right)^2
% \left( \frac{r_{\text{c}}}{\text{au}} \right)^{q_2-1}
% \left( \frac{t}{t_{\text{pff}}} \right)^{\frac{2}{3}(q_2-1)}
% \left( 1 + \frac{t}{t_{\text{pff}}} \right)^{-\gamma} 
% M_{\odot} \text{yr}^{-1}.
% \label{eq:pebble_mass_flux}
% \end{multline}
%
\begin{multline}
    \dot{M}_F = 2.1 \times 10^{-2} 
    \left( \frac{\ln \frac{R_{\text{peb}}}{R_0}}{\ln 10^4} \right)^{-1}
    \alpha_3^{-1} 
    %\dot{M}_{\text{acc8}} \\
    \frac{\dot{M}_{\text{acc}}}{10^{-8} M_{\odot}}\\
    \left( \frac{\Sigma_{\text{pg},0}}{\Sigma_{{\rm pg},\odot}} \right)^2
    \left( \frac{r_{\text{c}}}{\text{au}} \right)^{q-1} 
    \left( \frac{t}{t_{\text{pff}}} \right)^{\frac{2}{3}(q-1)}
    \left( 1 + \frac{t}{t_{\text{pff}}} \right)^{-\gamma} 
    M_{\odot} \text{yr}^{-1} \, ,
    \label{eq:pebble_mass_flux}
\end{multline}
where \(R_0\) and \(R_{\text{peb}}\) represent the initial size of dust particles and the eventual size of pebbles at which they commence migration, 
respectively. For this work, we have chosen typical values of \(R_{\text{peb}} \approx 10\,{\rm cm}\) and \(R_0 \approx 1\,{\rm \mu m}\). 
%in the aforementioned equation. 
%And \(T_2\) denotes a characteristic disc temperature within the irradiation region, with a value of \(T_2 = 150\) K being the standard in our default model. 
We replaced \(\dot{M}_*\) in \cite{Matsumura2021} with \(\dot{M}_{\rm acc}\) because the disc's local accretion rate is no longer the same as $\dot{M_*}$. %constant in disc radius. 
% Here, we also define the parameter
%
% \begin{equation}
% \dot{M}_{\text{acc8}} = \frac{\dot{M}_{\text{acc}}}{10^{-8} M_{\odot}}
% \end{equation}
% to characterise the radial dependence resulting from the disc wind. And 
$\Sigma_{\text{pg},0}$ is the initial pebble-to-gas surface mass density ratio and it is related to the stellar metallicity through
\begin{equation}
\frac{\Sigma_{\text{pg}}}{\Sigma_{\text{pg},\odot}} \approx 10^{\left[\mathrm{Fe}/\mathrm{H}\right]} \, ,
\label{eq:metallicity}
\end{equation}
where $\Sigma_{\text{pg},\odot} = 0.01$ is the value for the solar system.
%where $\Sigma_{\text{dg}} = \frac{\Sigma_{\text{d}}}{\Sigma_{\text{g}}}$, and $\Sigma_{\text{dg},\odot} = 0.01$ is the value for the solar system. We assume $\Sigma_{\text{d}} = \Sigma_{\text{p}}$ at the pebble formation front. 
For our simulations, we have considered five different metallicities: $\left[\mathrm{Fe}/\mathrm{H}\right] = (-0.5, -0.3, 0, 0.3, 0.5)$\footnote{In this work, we assume that the wind mass loss does not change the dust-to-gas ratio of the disc.  If the wind mass loss increases the dust-to-gas ratio, the planets would grow in a more metal-rich environment, which could affect the formation outcome.  We will leave this investigation for a future study.}.
%The exponent $q_2$ in the temperature profile $T \propto r^{-q_2}$ is adopted as $\frac{3}{7}$, based on the irradiation regime \citep{Ida2016}. 
%The exponent $q$ in the temperature profile $T \propto r^{-q}$ is $q=\frac{3}{7}$ in the irradiation regime \citep{Ida2016}. \textcolor{black}{i moved this line to section 2.1.2, remove this line?} 
The timescale for the pebble formation front to reach the outer disc radius, $t_{\text{pf}}$, is also calculated as in \cite{Matsumura2021} as
\begin{equation}
%t_{pf} \approx \ln \left( \frac{R_{peb}}{R_0} \right) \cdot \frac{200}{\sqrt{3\pi^3}} \left( \frac{\Sigma_{pg,0}}{0.01} \right)^{-1} \left( \frac{M_*}{M_{\odot}} \right)^{-1/2} \left( \frac{r_c}{\text{au}} \right)^{3/2} \text{yr}
t_{pf} \approx 21\ln \left( \frac{R_{peb}}{R_0} \right) \left( \frac{\Sigma_{pg,0}}{0.01} \right)^{-1} \left( \frac{M_*}{M_{\odot}} \right)^{-1/2} \left( \frac{r_c}{\text{au}} \right)^{3/2} \text{yr} \, .
\label{eq:pebble formation timesacle}
\end{equation}

Similar to \(\dot{M}_{\text{acc}}\), the mass flux \(\dot{M}_F\) is also a function of both time \(t\) and radial distance \(r\). 
Since \(\dot{M}_F\) is proportional to \(\dot{M}_{\text{acc}}\), the pebble mass flux is higher in the outer disc. 
%and thus pebble accretion rate is higher in the outer disc.
%the increased pebble mass flux significantly enhances the pebble accretion rate at the outer disc.

%The pebble accretion rate is influenced by the pebble accretion efficiency, denoted as \( \epsilon \). 
%This efficiency represents the fraction of pebbles that are effectively accreted onto a growing planetary core. 
%Thus, the pebble accretion rate can be expressed as 
Only a fraction of this pebble mass flux is accreted onto a protoplanet, and thus the pebble accretion rate 
can be expressed as follows by using the pebble accretion efficiency $\epsilon$:
\begin{equation}
\dot{M}_p = \epsilon \dot{M}_F \, .
\label{eq:pebble_accretion}
\end{equation}

We adopted the expression for pebble accretion efficiency $\epsilon$ as developed by \cite{Ida2016}:
\begin{equation}
\epsilon = \min \left( \frac{C \zeta^{-1} \chi \hat{b}^2}{4 \sqrt{2 \pi} \tau_s \hat{h}_p} \left( 1 + \frac{3\hat{b}}{2 \chi \eta} \right), 1 \right) \, , 
\label{eq:pebble_accretion_efficiency}
\end{equation}
where the efficiency is capped at unity \textcolor{black}{and $\eta=0.5\,\hat{h}_g^2\left|\frac{d\ln P}{d\ln r}\right|$ with $P$ being the pressure}.  %, as the term in parentheses indicates. 
The variables \( \chi\) and \( \zeta \) are functions of the Stokes number \( \tau_s \), and are defined as follows:  
\begin{equation}
\chi = \sqrt{\frac{1 + 4\tau_s^2}{1 + \tau_s^2}} \, ,
\label{eq:chi}
\end{equation}
\begin{equation}
\zeta = \frac{1}{1 + \tau_s^2} \, .
\label{eq:zeta}
\end{equation}

The pebble aspect ratio is represented by \( \hat{h}_p \) and is the ratio of the pebble scale height \( h_p \) 
to the radial distance \( r \), where \( h_p \) is approximated by
\begin{equation}
h_p \approx \left( 1 + \frac{\tau_s}{\alpha_{\text{SS}}} \right)^{-1/2} h_g.
\label{eq:hp}
\end{equation}

The parameter \( \hat{b} \), related to the impact parameter \( b \) by \( \hat{b} = \frac{b}{r} \), 
describes the interaction zone for a pebble approaching a protoplanet with mass \( M_p \). 
The impact parameter \( b \) is estimated as,
\begin{equation}
b \approx \min \left( \sqrt{\frac{3\tau_s^{1/3} R_H}{\chi \eta r}}, 1 \right) \times 2\kappa \tau_s^{1/3} R_H \, ,
\end{equation}
where the terms inside the minimum function refer to the Bondi and Hill accretion regimes, respectively. 
The Hill radius \( R_H \) is given by \( R_H = \left( \frac{M_p}{3M_*} \right)^{1/3}r \), and \( \kappa \) 
is the reduction factor for \( b \) when \( \tau_s \gg 1 \). 
The expression is derived by \cite{OrmelKobayashi2012} as
\begin{equation}
\kappa = \exp \left( -\left( \frac{\tau_s}{\min(2, \tau_s^*)} \right)^{0.65} \right) \, ,
\end{equation}
with \( \tau_s^* \) being \(\tau_s^* = 4 \left( \frac{M_p}{M_*} \right) \eta^{-3}\). 
Lastly, \( C \), is given by
\begin{equation}
C = \min \left( \sqrt{\frac{8}{\pi}} \frac{h_p}{b} , 1 \right) \, ,
\end{equation}
which correspond to 2D and 3D accretion conditions, respectively.

%\subsubsection{Halt Pebble Accretion}
\subsubsection{Pebble \textcolor{black}{isolation mass}}
\label{subsubsec:halt-pebble-accretion}

Pebble isolation mass (PIM) is a critical mass in the growth of planetary cores. 
This stage is marked by a significant alteration in the gas surface density profile caused 
by the gravitational influence of the growing protoplanet. 
The alteration is such that it reverses the pressure gradient in the vicinity of the protoplanet. 
Consequently, instead of experiencing headwind that facilitates their inward drift, pebbles encounter a tailwind, 
effectively halting the flow of pebbles \citep[e.g.,][]{Johansen2017,Ormel2017} 
%This change effectively halts the flow of pebbles to the protoplanet. 
In this work, the PIM from \cite{Ataiee2018} is adopted as follows\textcolor{black}{:}
\begin{align}
    M_{\text{iso}} &\approx \hat{h}_g^3 \sqrt{37.3\alpha_{\text{SS}} + 0.01} \nonumber \\ 
    &\quad \times \left[ 1 + 0.2 \left( \frac{\sqrt{\alpha_{\text{SS}}}}{\hat{h}_g} \sqrt{\frac{1}{\tau_s^2} + 4} \right)^{0.7} \right] \times M_*
    \label{eq:pebble-isolation-mass}
\end{align}
%
%In \cite{Ataiee2018}, it was found that the surface mass density of the gap created by protoplanets reaching PIM can decrease by up to 20\%. 
In \cite{Ataiee2018}, it was found that protoplanets reaching PIM opens a gap and the surface mass density there 
decreases by up to 20\%.
On the other hand, \cite{Johansen2019} identified the gap transition mass that corresponds to a 50\% gap, 
where migration switches from type I to type II:
\begin{equation}
M_{\text{trans}} = \left( \frac{\alpha_{\text{SS}}}{0.04} \hat{h}_g^5 \right)^{1/2} M_* \, .
\label{eq:transition mass}
\end{equation}
However, as indicated in Figure~3 in \cite{Matsumura2021}, the planetary mass to reach a $50\%$ gap becomes 
smaller than that to reach a $20\%$ gap in the outer part of the disc, which is 
counter-intuitive.
%However, \citep{Matsumura2021} indicated that the transition mass \( M_{\text{trans}} \), which signifies the shift from type I to type II migration, corresponds to a gap depth of 50\% and is defined as:
%
%\begin{equation}
%M_{\text{trans}} = \left( \frac{\alpha_{\text{SS}}}{0.04} \hat{h}_g^5 \right)^{1/2} M_*
%\label{eq:transition mass}
%\end{equation}
%
%This mass is lower than the PIM in the outer disc. The development of Equation \ref{eq:transition mass} implies that protoplanets in the outer disc switch to Type II migration without reaching PIM, which is counter-intuitive.
%
Therefore, in our simulations, pebble accretion ceases once the planetary mass reaches \(\min(M_{\text{iso}}, M_{\text{trans}})\).
%, and gas accretion commences immediately.

\subsection{Gas \textcolor{black}{envelope accretion}}
\label{subsec:gas-envelope-accretion}
%
%When a massive core forms with a low solid accretion rate, gas can contract and form an envelope. 
The critical core mass marks the point beyond which the hydrostatic equilibrium breaks down for the gas envelope\textcolor{black}{,} 
and thus the envelope contraction rather than the mass accretion is necessary to support the envelope \citep{Ikoma2000}. 
This envelope contraction leads to rapid gas accretion\textcolor{black}{,} and thus the critical core mass is often used as the starting point of 
gas accretion in planet formation models. 
We adopted a fitting formula of a critical core mass as a function of pebble accretion rate derived by \cite{OgiharaHori2020}:
%We have adopted \cite{OgiharaHori2020}'s fitting function to estimate the critical core mass 
%beyond which the hydrostatic equilibrium breaks down and thus the   
%to start gas accretion as follows:
%
\begin{equation}
    M_{\text{crit}} = 13 \left( \frac{\dot{M}_p}{10^{-6} M_{\oplus} \, \text{yr}^{-1}} \right)^{0.23} M_{\oplus} \, .
    \label{eq:critical-mass-gas-accretion}
\end{equation}
The above equation suggests that gas accretion commences when the mass accretion rate $\dot{M}_p$ becomes small enough, 
either due to the attainment of the PIM or the depletion of the pebble flux.

Once this condition is met, the gas accretion onto a protoplanet is constrained 
by both the rate at which a protoplanet can accumulate gas and the rate at which a disc can deliver gas to the protoplanet \citep{Ida2018}:
\begin{equation}
%\frac{dM_p}{dt} \approx \min \left[ \frac{dM_{p,\text{KH}}}{dt}, \dot{M}_{\text{acc}}, f_{\text{local}} \dot{M}_{\text{acc}} \right].
\dot{M}_p \approx \min \left[ \dot{M}_{\text{KH}}, \dot{M}_{\text{acc}}, \dot{M}_{\rm TT16} \right].
\end{equation}
%
%The process of gas accretion onto a protoplanetary core, particularly during the Kelvin-Helmholtz contraction phase is characterised 
%by a gradual increase in the core's gravitational pull, leading to the accretion of gas from the surrounding protoplanetary disc. 
%The rate of gas accretion during this phase is given by the following equation, as described by \citep{OgiharaHori2020}:
Initially, a protoplanet core accretes gas according to the Kelvin-Helmholtz contraction rate \citep{Hori2010}:
\begin{equation}
    \dot{M}_{\text{KH}} = 10^{-8} \left( \frac{M_p}{M_{\oplus}} \right)^{3.5} M_{\oplus} \text{yr}^{-1} \, .
    \label{eq:kelvin-helmholtz-accretion}
\end{equation}
%
%In Equation \ref{eq:kelvin-helmholtz-accretion}, \( \dot{M}_{\text{KH}} \) represents the Kelvin-Helmholtz gas accretion rate, and 
%\( M_{\text{p}} \) denotes the mass of the protoplanetary core. 
%The Kelvin-Helmholtz contraction phase is crucial for the significant formation of a gas envelope around the core, marking the transition from a predominantly solid core to a gas-rich giant planet. 
%The underlying physics of this process involves a balance between the core's gravitational pull and the thermal pressure of the accreted gas, with the rate of accretion increasing sharply as the core mass grows. However, 
%
As the mass of the protoplanets' gaseous envelope increases, the gas accretion rate is determined either by $\dot{M}_{\text{acc}}$
or by $\dot{M}_{\rm TT16} = f_{\text{local}} \dot{M}_{\text{acc}}$ \textcolor{black}{\citep{Tanigawa2016}}, where $f_{\text{local}}$ is the accretion efficiency defined as
\begin{equation}
    f_{\text{local}} \approx \frac{0.031 \hat{h}^{-4} \left( \frac{M_p}{M_*} \right)^{4/3}}{1 + 0.04K} \alpha_{\rm total}^{-1} \, ,    %\tilde{\alpha}^{-1}.
    \label{eq:f_local}
\end{equation}
with K being the gap depth factor defined by \cite{Kanagawa2015} as
\begin{equation}
    K = \left( \frac{M_p}{M_*} \right)^2 \left( \frac{h}{r} \right)^{-5} \alpha_{\text{SS}}^{-1} \, .
\end{equation}
When $\dot{M}_{\rm TT16} < \dot{M}_{\text{acc}}$, \textcolor{black}{a} sufficient amount of gas is supplied to a protoplanet\textcolor{black}{,} and 
thus the gas accretion rate is regulated by how quickly the protoplanet can accrete gas.  
When $\dot{M}_{\rm TT16} > \dot{M}_{\text{acc}}$, on the other hand, the gas accretion rate is 
determined by how quickly the disc can supply gas to the protoplanet.
%
%When $\dot{M}_{\rm TT16} < \dot{M}_{\text{acc}}$, the gas accretion is regulated by the  
%by how quickly the disc can supply gas to the vicinity of the protoplanets \citep{Ida2018}. 
%
% Thus, during this stage, the gas accretion rate is determined by \( \dot{M}_{\text{acc}} \). 
% Additionally, \cite{TanigawaTanaka2016} proposed that when \( M_p > 10 M_{\text{J}} \), 
% the formation of a gap around the growing planets would reduce the gas supply rate. 
% Therefore, a gas accretion reduction factor \( f_{\text{local}} \) should also be included. 
%
%
% We have adopted the formula from \cite{Ida2018}:
% %
% \begin{equation}
% f_{\text{local}} \approx \frac{0.031 \hat{h}^{-4} \left( \frac{M_p}{M_*} \right)^{4/3}}{1 + 0.04K} \alpha_{\rm total}^{-1} \, ,    %\tilde{\alpha}^{-1}.
% \label{eq:f_local}
% \end{equation}
% %
% with K being the gap depth factor defined as \citep{Kanagawa2015},
% %
% \begin{equation}
% K = \left( \frac{M_p}{M_*} \right)^2 \left( \frac{h}{r} \right)^{-5} \alpha_{\text{ss}}^{-1} \, .
% \end{equation}
%
%Eventually, as the mass of the gas envelope \( M_{\text{env}} \) becomes comparable to the core mass \( M_{\text{core}} \), 
%the protoplanets initiate runaway gas accretion \citep{Pollack1996}. 
%In a later section, we demonstrate that at this stage in the wind-driven disc, the protoplanets accrete gas almost in-situ.

\subsection{Planet migration}
\label{subsec: Planet migration}

When the masses of protoplanets are low, they cannot open up a gap and thus undergo type I migration with a timescale \textcolor{black}{of} \( \tau_{\text{mig1}} \). 
We adopted the equation from \cite{Ida2018}:
\begin{equation}
\tau_{\text{mig1}} \approx \frac{1}{2c} \left( \frac{M_p}{M_*} \right)^{-1} \left( \frac{\Sigma r^2}{M_*} \right)^{-1} \hat{h}^{2} \Omega^{-1} \, .
\label{eq: mig1}
\end{equation}
%
%where \( \Omega \) is the Keplerian frequency. 
As shown by \cite{Kanagawa2018}, the parameter \( c \) typically ranges between 1 and 3, 
aligning well with the isothermal formula for type I migration outlined by \cite{Tanaka2002}. 
In this work, we adopt \( c = 2 \).

As the mass of the protoplanets continually increases, they can open a gap and the migration switches to type II migration.
\cite{Kanagawa2018} proposed the timescale for type II migration is essentially the same as the type I migration timescale, 
but with the gas surface density reduced in the gap. The timescale for type II migration is expressed as, 
%
%\begin{align}
%\tau_{\text{mig2}} &\approx (1 + 0.04K) \tau_{\text{mig1}} \\
%&\approx \frac{1}{2c} \left[ 1 + 0.04 \left( \frac{M_p}{M_*} \right)^2 \hat{h}^{-5} \alpha_{\text{ss}}^{-1} \right] \\
%&\quad \times \left( \frac{M_p}{M_*} \right)^{-1} \left( \frac{\Sigma r^2}{M_*} \right)^{-1} \hat{h}^2 \Omega^{-1}.
%\end{align}
\begin{equation}
    \tau_{\text{mig2}} \approx (1 + 0.04K) \tau_{\text{mig1}} \, ,
    \label{eq: mig2}
\end{equation}
where migration swithces from type I to type II when $K=1/0.04$ \citep{Johansen2019} or 
when the planetary mass corresponds to Equation~\ref{eq:transition mass}.

\subsection{Initial condition}
\label{subsec: ICs}

In this study, nine disc models with different initial masses and \(\alpha_{\text{total}}\) values are tested around a Sun-like star, 
as summarised in Table \ref{tab:disc_parameters} (also see Figure~\ref{fig:stellar_acc_rate}). 
%The temporal evolution of the stellar accretion rates of these discs is presented in Fig. \ref{fig:stellar_acc_rate}. 
Five metallicities, \([ \text{Fe/H} ] = -0.5, -0.3, 0, 0.3,\) and \(0.5\) 
and three magnetic lever arms, \(\lambda = 1.6, 3,\) and \(17\), are considered for each disc. 
The disc 5 with $\lambda = 3$ and [Fe/H] = 0 is selected as the fiducial case. 

In each disc, 40 non-interacting embryos are logarithmically spaced between 1 and 100 au to track planet formation across the disc. 
\textcolor{black}{Although a characteristic mass of a planetesimal formed via streaming instability is about a Ceres mass \citep[$\sim 1.6\times10^{-4}\,M_{\oplus}$,][]{Johansen2015}, we have chosen the initial mass of an embryo to be \(0.01\,M_{\oplus}\) for simplicity.
The choice of a constant initial embryo mass across the disc is rather arbitrary, but this assumption ensures that the initial core masses are the same not only at different disc radii but also for different disc models. 
Furthermore, by adopting the initial masses that are two orders of magnitude higher than a Ceres mass, we are implicitly assuming that a protoplanet grows upto the pebble onset mass via planetesimal-planetesimal collisions.
\cite{Johansen2017} showed that the pebble accretion onset mass is $\sim10^{-3}-10^{-2}\,M_{\oplus}$ in $0.4-4\,$au, and reaches up to \(\sim 10^{-1}\,M_{\oplus}\) in the outer disc region (40 to $100$~au).  \cite{Lau2022}, on the other hand, studied the formation and growth of planetary cores in a pressure bump and found the onset mass for pebble accretion can be reduced to \(\sim 10^{-4} M_{\oplus}\). In Section~\ref{sec:Discussion}, we briefly discuss the effects of radially dependent initial core masses.}

In all simulations, planet formation ceases either when all the gas is depleted or, for the particularly long-lived discs 7, 8, and 9, 
when the discs reach their artificially set lifetimes of 100\,Myr
\footnote{This long disc lifetime does not affect the overall outcome of planet formation for most cases. 
For $\lambda=3$, the exception is Disc 9 with ${\rm [Fe/H]}=0$ (see Figure~\ref{fig:meta_0_lambda_3}) where 
some planets became giant planets within $20-30$~Myr.
Even later gas accretion was observed for $\lambda=1.6$ with ${\rm [Fe/H]}=0.5$ and 0.3 in Disc 9 and with ${\rm [Fe/H]}=-0.3$ in Disc 8, 
but we keep these results for the analysis because they don't change the overall results.}.
%where some planets became giants at $\sim100\,$Myr (see Figures~\ref{fig:meta_0_lambda_3} for example).
%(see Figures~\ref{fig:meta_0_lambda_3} and \ref{fig:meta_0_lambda_17} for example).  
%For Discs 7 and 8, the planetary masses did not change more than a factor of \textcolor{black}{****} between $10\,$Myr and and $100\,$Myr.
%\textcolor{black}{Any comments on different metallicities and/or lambda??}}.

\begin{table}[ht]
\centering
\begin{tabular}{ccc}
\hline
\textbf{Disc} & \( M_{D,0} (M_{\odot}) \) & \( \alpha_{\text{total}}\) \\ \hline 
%(= \alpha_{\rm SS} + \alpha_{\rm WD}) \) \\ \hline
disc1 & 0.2   & \( 7.4 \times 10^{-2} \) \\
disc2 & 0.06  & \( 7.4 \times 10^{-2} \) \\
disc3 & 0.02  & \( 7.4 \times 10^{-2} \) \\
disc4 & 0.2   & \( 7.4 \times 10^{-3} \) \\
disc5 & 0.06  & \( 7.4 \times 10^{-3} \) \\
disc6 & 0.02  & \( 7.4 \times 10^{-3} \) \\
disc7 & 0.2   & \( 7.4 \times 10^{-4} \) \\
disc8 & 0.06  & \( 7.4 \times 10^{-4} \) \\
disc9 & 0.02  & \( 7.4 \times 10^{-4} \) \\
\hline
\end{tabular}
\caption{\textcolor{black}{Initial} disc mass $M_{D,0}$ and the total value of the parameter $\alpha_{\rm total}$ of nine disc models.}
\label{tab:disc_parameters}
\end{table}

\section{Results}
\label{sec:Results}

\subsection{Growth tracks of planets in wind-driven and viscously evolving discs}\label{subsec:growth_tracks}

% In all simulations, planet formation ceases either when all the gas is depleted or, for the particularly long-lived discs 7, 8, and 9, 
% when the discs reach their artificially set lifetimes of 100\,Myr
% \footnote{This long disc lifetime does not affect the overall outcome of planet formation except for Disc 9, 
% where some planets became giants at $\sim100\,$Myr (see Figures~\ref{fig:meta_0_lambda_3} and \ref{fig:meta_0_lambda_17} for example).  
% For Discs 7 and 8, the planetary masses did not change more than a factor of **** between $10\,$Myr and and $100\,$Myr.}
%
% Planets that migrated to the inner edge of the disc (0.01\,au) are considered lost; 
% however, their growth trajectories are still depicted in the figures for consistency.

\begin{figure*}[ht]
\centering
\includegraphics[width=\textwidth]{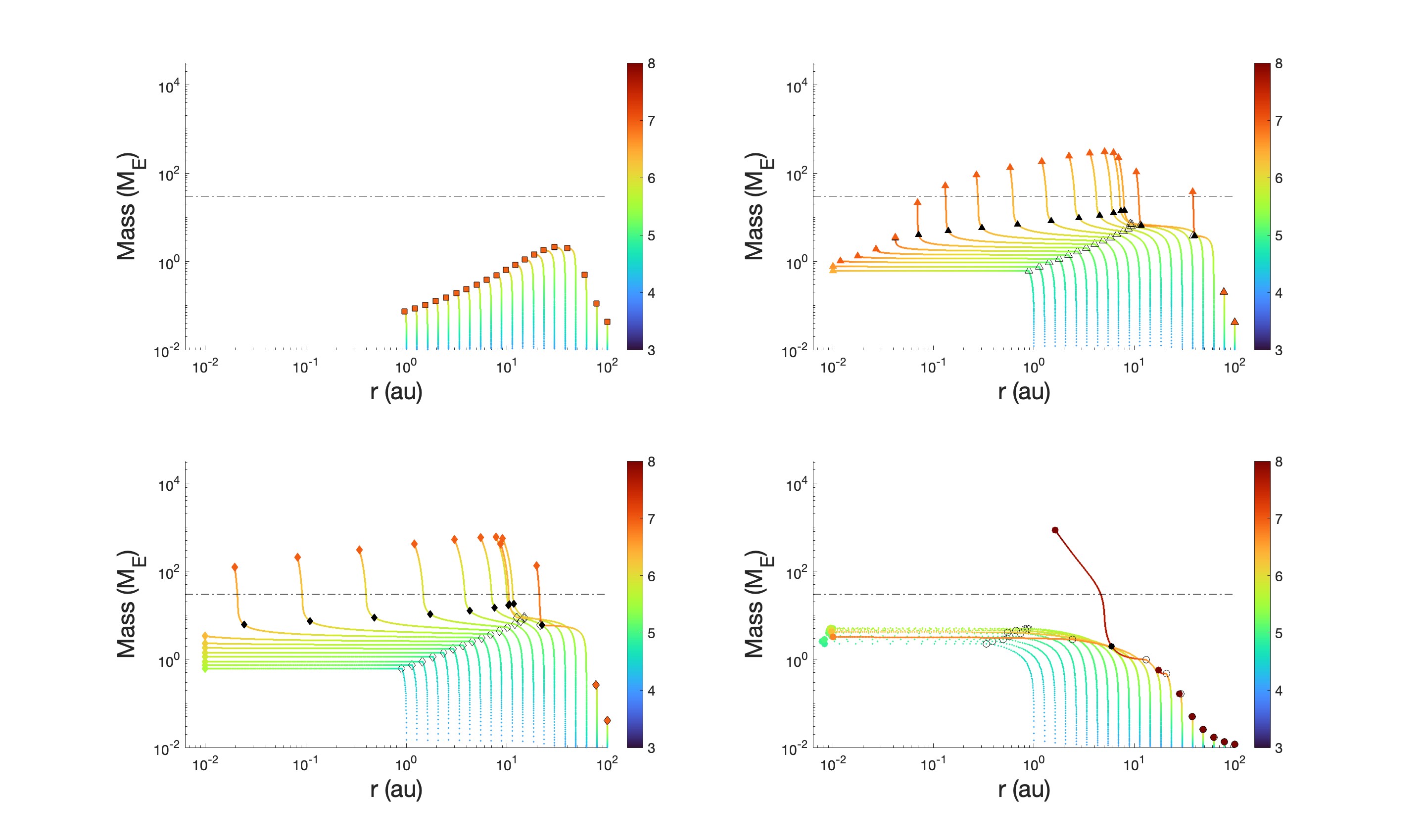}
\caption{Growth tracks of Disc 5 with \textcolor{black}{[Fe/H]=0}. Top left, top right, bottom left, and bottom right panels correspond to 
WD discs with $\lambda = 1.6$, 3, 17, and VE disc, respectively. 
Black open diamonds represent the core masses of the planets, while black solid diamonds show where \(M_{\text{core}} = M_{\text{env}}\).
The black dash-dotted lines denote a total mass of \(30 M_{\oplus}\).
}
\label{fig:disc_5_meta_0_lambda_3}
\end{figure*}

% \begin{figure*}[ht]
% \centering
% \includegraphics[width=\textwidth]{fig/disc5_meta_0_lambda_17.jpg}
% \caption{Growth tracks of Disc 5 with 0 metallicity and $\lambda = 17$}
% \label{fig:disc_5_meta_0_lambda_17}
% \end{figure*}

% \begin{figure*}[ht]
% \centering
% \includegraphics[width=\textwidth]{fig/disc_5_meta_0_vis.jpg}
% \caption{Growth tracks of Disc 5 with zero metallicity and purely viscously evolving.}
% \label{fig:disc_5_meta_0_vis}
% \end{figure*}

%Firstly, Fig.\ref{fig:disc_5_meta_0_lambda_3}, Fig.\ref{fig:disc_5_meta_0_lambda_17}, and Fig.\ref{fig:disc_5_meta_0_vis} 
%show the growth tracks of the planets in the fiducial disc, the same disc with $\lambda = 17$, and the same disc but purely viscously evolving. 
First, we will present the general planet formation outcomes for various wind efficiencies of WD discs as well as a VE disc.
Figure~\ref{fig:disc_5_meta_0_lambda_3} shows the growth tracks of the planets in Disc 5 with \textcolor{black}{the solar metallicity [Fe/H]=0}
for $\lambda=1.6$ (upper left), $\lambda=3$ (upper right), $\lambda=17$ (lower left), and a VE disc (lower right).
%
%
%These figures illustrate the final planetary mass at the end of the simulation for 20 embryos logarithmically located between 1 and 100 au 
%instead of the default 40 embryos for clarity. 
These figures illustrate the mass growth tracks of logarithmically-separated, 20 non-interacting embryos instead of the default 40 embryos 
for clarity.  
The colour gradient indicates the \(\log_{10}\) of the time taken to reach a certain mass. 
Black hollow symbols represent the critical core masses of the planets where they start gas accretion, 
while black solid symbols show where \(M_{\text{core}} = M_{\text{env}}\)\textcolor{black}{,} which approximately corresponds to
the onset of rapid gas accretion \citep{Pollack1996}. 
%when the planets trigger rapid gas accretion \citep{Pollack1996}. 
For the WD cases, the hollow symbols usually correspond to PIMs, but there are also cases in which cores start gas accretion before reaching PIMs. 
%(\(M_{\text{core}} \approx M_{\text{env}}\)). 
The black dash-dotted lines denote a total mass of \(30 M_{\oplus}\). Planets above these lines are considered giant planets in this work.
Planets that migrated to the inner edge of the disc (0.01\,au) are assumed lost; 
however, their growth trajectories are still depicted in the figures for consistency.

%Each panel corresponds to the same initial disc mass and the same $\alpha_{\rm total}$, but a different wind mass-loss rate 
%and thus a different stellar mass accretion rate as shown in Figure~\ref{fig:stellar_acc_rate} (see magenta lines).
%As seen from magenta lines for Disc 5 in Figure~\ref{fig:stellar_acc_rate}, t
The four cases shown in Figure~\ref{fig:disc_5_meta_0_lambda_3} have the same initial disc mass and the same $\alpha_{\rm total}$, but different stellar mass accretion rates 
due to various wind mass-loss rates (see magenta lines for Disc 5 in Figure~\ref{fig:stellar_acc_rate}). 
This leads to very different outcomes of planet formation. 

First, the VE disc case has few surviving planets due to rapid type I migration, which 
agrees with the outcome of previous studies \citep[e.g.,][]{Matsumura2017}.  
This is primarily because the assumed $\alpha_{\rm total}=\alpha_{\rm SS}=7.3\times10^{-3}$ is too high to allow the survival of planetary cores.  
One giant planet survived at $\sim2\,$au, but the initial core was located very far from the central star ($\sim20\,$au), which 
also agrees with the trend observed by previous pebble accretion studies with a comparable disc's viscosity \citep[e.g.,][]{Bitsch2019}.

The $\lambda=17$ case has the lowest wind mass-loss rate and there is only a weak radial dependence of the mass accretion rate 
(see Figure~\ref{fig:local_acc_rate}).
This case has the similar initial surface mass density profile to the VE disc (see Figure~\ref{fig:sigma0disc5}) 
as well as the comparable stellar mass accretion rate up to $\sim10\,$Myr (see Figure~\ref{fig:stellar_acc_rate}).
However, planet formation outcome of the $\lambda=17$ disc is quite different from the VE disc, because the disc turbulence parameter 
is much smaller in the WD disc ($\alpha_{\rm SS}=10^{-4}$).  
A smaller $\alpha_{\rm SS}$ leads to a thinner pebble layer (see Equation~\ref{eq:hp}), which accelerates the pebble accretion.
Once the PIMs are reached, protoplanetary cores accrete gas rapidly and become giant planets.
The migration appears to slow down as the rapid gas accretion occurs because type I migration becomes faster with mass while type II 
migration becomes slower (see Equations~\ref{eq: mig1} and \ref{eq: mig2}). 

As it can be seen from the number of lost cores, planet formation with $\lambda=17$ is much faster than the other two $\lambda$ cases, 
and only giant planets have survived. 
Among WD discs with different $\lambda$ values, $\alpha_{\rm total}$, $\alpha_{\rm SS}$, and $\alpha_{\rm DW}$ are the same\textcolor{black}{,} and 
thus the pebble scale height $h_p$, the pebble accretion efficiency $\epsilon$, 
and the pebble isolation mass $M_{\rm iso}$ are independent of $\lambda$.  
The difference in planet formation rates among WD discs arises from the difference 
in the pebble mass flux $\dot{M}_F$ that is proportional to the local disc accretion rate $\dot{M}_{\rm acc}$ (see Equation~\ref{eq:pebble_mass_flux}).  
The discs with $\lambda=17$ have the least wind mass-loss rate and thus have the highest local disc accretion rate.

For $\lambda=17$, giant planets are formed out of cores that are initially at $\sim7\,$au or beyond, 
and various types of giant planets (i.e., hot, warm, and cold Jupiters) are seen from $\sim0.02\,$au to $\sim30\,$au.  
However, giant planets with wide orbital radii (beyond a few tens of au) were not formed because 
pebble accretion is slow there for our initial core masses. 
%\textcolor{black}{Do the outer cores have the pebble onset mass above 1e-2??}
%\textcolor{black}{Yes, the pebble onset mass increase up to~1e-1 in the outer disc region (40-100au).}

Here, we also have in-situ formation of hot Jupiters with semimajor axes $\lesssim0.1\,$au as well as warm Jupiters with semimajor axes of $\sim0.1-1\,$au. 
Such formation of giant planets near the central star might be limited due to the recycling of the planetary atmosphere, where the flow of gas from the 
disc injecting a higher entropy gas into the planetary atmosphere and slowing down the atmospheric cooling and further gas accretion \citep[e.g.,][]{Ormel2015b,Ali-Dib2020}.
Since we have not taken this effect into account, it is possible that we are overestimating the number of hot Jupiters.  
For warm Jupiters, however, \cite{Savignac2023arXiv} recently showed that such a recycling effect is not very strong at $0.1\,$au and weaker outside it, 
making the in-situ formation of WJs and potentially some HJs plausible.  
Even when in-situ formation of hot Jupiters is not possible, a disc \textcolor{black}{similar to} this may still be able to form such planets. 
If multiple giant planets are formed beyond $\sim1\,$au, they could experience dynamical instabilities \citep[e.g.,][]{Chatterjee2008,Juric2008,Matsumura2010a}.
When the pericentre of a planetary orbit becomes small enough, a hot Jupiter or a warm Jupiter can be formed via the tidal circularisation.

The $\lambda=3$ case has the intermediate wind mass-loss rate and has resulted in formation of both low-mass and high-mass planets.  
This kind of mass distribution was not observed in \cite{Matsumura2021}, where we assumed ``two-alpha'' disc models, because 
the local disc mass flux was constant in radii.
Some cores are lost to the central star due to rapid type I migration, but all planets formed from cores that are initially beyond $\sim1.5\,$au survived.
There is also a general mass gradient inside out. 
The surviving low-mass planets have small-orbital radii of $<0.1\,$au while the surviving giant planets have orbital radii ranging from $\sim0.1\,$au to several tens of au.  
These giant planets are formed from cores with initial radii of $\gtrsim5\,$au, and their masses peak for cores with initial radii of $\sim20-30\,$au. 
A disc \textcolor{black}{such as} this may lead to formation of either a warm Jupiter with a low-mass inner companion or a short-period super Earth accompanied by a cold Jupiter. 

The $\lambda=1.6$ case has the largest wind mass-loss rate and thus has the steepest radial dependence of the disc mass accretion rate 
(see Figure~\ref{fig:local_acc_rate}).
Thus, the dust mass flux also sharply decreases \textcolor{black}{towards} the central star and planet formation is slow. 
As a result, the disc only formed low-mass planets that did not reach PIMs. 

\begin{figure*}[ht]
\centering
\includegraphics[width=\textwidth]{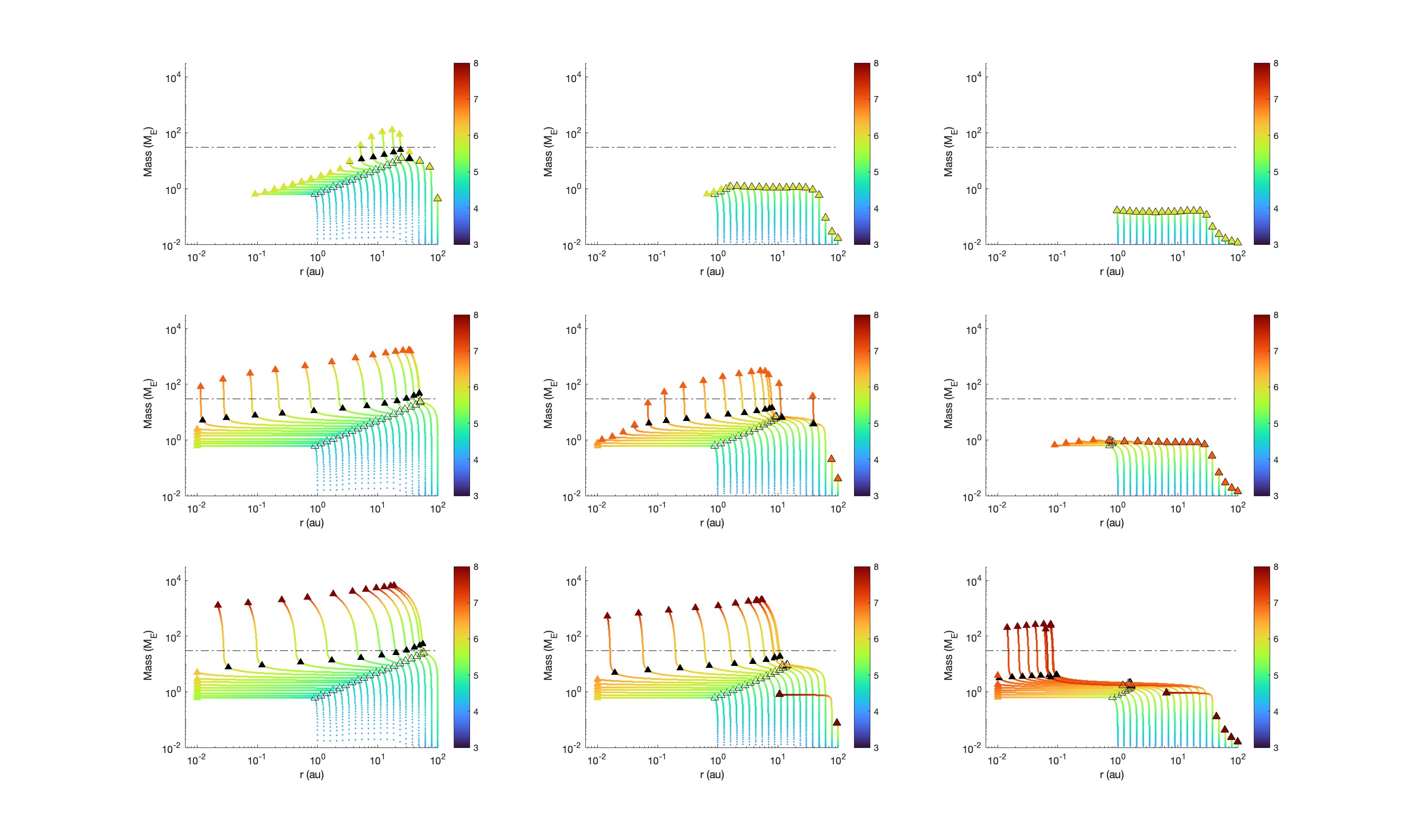}
\caption{Outcomes and evolution trajectories of single-core planet formation in \textcolor{black}{nine} wind-driven discs with \textcolor{black}{the solar metallicity [Fe/H]=0}, $\lambda = 3$. The first row, from left to right, corresponds to discs 1 to 3; the second row represents discs 4 to 6; and the third row denotes discs 7 to 9.}
\label{fig:meta_0_lambda_3}
\end{figure*}

These three types of planet formation trends are not unique to each $\lambda$.  As seen in Figure~\ref{fig:meta_0_lambda_3}, 
all types of planet formation (i.e., only low-mass, both low- and high-mass, and only high-mass planets) arise for 
appropriate combinations of the mass, lifetime, and metallicity of a disc for each $\lambda$.  
However, as we discuss in Section~\ref{subsec:lambda_mass}, a difference in $\lambda$ does result in a difference 
in the overall planetary mass distribution of a system.  \\

%Additionally, we also show here the growth tracks of $\lambda = 3$ and $17$ for nine disc models summarised in Table~\ref{tab:disc_parameters}.
Figure~\ref{fig:meta_0_lambda_3} shows similar planet formation outcomes for Discs~1-9 with \textcolor{black}{[Fe/H]=0} and $\lambda=3$.
Here, the initial disc mass decreases from left to right (i.e., $0.2\,M_{\odot}$, $0.06\,M_{\odot}$, and $0.02\,M_{\odot}$) 
while the disc's lifetime increases from top to bottom (i.e., $\alpha_{\rm total}=7.4\times10^{-2}$, $7.4\times10^{-3}$, and $7.4\times10^{-4}$).
Overall, when disc's masses are low or lifetimes are short (e.g., upper-right panels, Discs~2, 3, and 6), planet formation is inefficient and only low-mass planets form. 
For higher (lower) $\lambda$ values, the overall trends are similar, but planet formation is faster (slower)\textcolor{black}{,} and 
thus more (less) massive planets are formed \textcolor{black}{closer in (further out).}

On the other hand, when disc's masses are high or lifetimes are long (e.g., lower-left panels, Discs~4, 7, and 8
\footnote{Disc~9 also forms giant planets, but only \textcolor{black}{towards} the very end of the allowed disc's lifetime of 100\,Myr.  
Since such a long lifetime is unrealistic, Disc~9 is likely to lead to only low-mass planets when taking account of the disc dissipation mechanism 
such as photoevaporation.}), planet formation is efficient and giant planets are formed across the entire disc.  
For higher (lower) $\lambda$ values, the overall trends are similar again, but more (less) cores are lost to the central star without 
becoming giant planets. 

Finally, for intermediate cases (e.g., Discs~1 and 5), we have found that both low-mass and high-mass planets can be formed in the same disc.  
\textcolor{black}{A system with such a jump in the mass distribution is characterised by two conditions: low-mass planets survive in the inner disc, and giant planets form further out. 
For both of these conditions to meet, we \textcolor{black}{have} found that the giant planet formation timescale needs to be roughly comparable to the disc accretion timescale ($t_{\rm PF}\sim t_{\text{acc},0}$), where the giant planet formation timescale measures the onset of rapid gas accretion ($M_{\rm core} \sim M_{\rm env}$).
Satisfying the first condition becomes somewhat easier in low $\lambda$ discs (e.g., $\lambda=1.6$), because the core survival rate becomes higher partly because it takes longer for inner cores ($\lesssim2\,$au) to reach the PIM and partly because type I migration slows due to the reduced surface mass density with a flatter profile.}
%The more challenging condition is the former, as rapid type I migration often results in the loss of cores. For low $\lambda$ discs (such as $\lambda=1.6$), the core survival rate can increase due to a decreased pebble mass flux in the inner disc, which delays the time for cores inside $\sim2\,$au to reach the PIM. Combined with high wind mass loss rate, this reduces the remaining surface mass density, consequently slowing down type I migration. 
%
% Such cases are observed in $\lambda = 1.6$ with high and intermediate disc mass discs 4 and 8, where cores initially located within $\sim2$ AU survive and form SEs in disc 4, and SEs plus HJs in disc 8. These SEs contribute to the larger mass diversity in these discs. In contrast, in the $\lambda = 3$ case, these cores are lost, leaving only giant planets across the entire disc.}

For higher (lower) $\lambda$ values, the overall trends are similar, though these intermediate cases \textcolor{black}{with a mass jump} happen for different disc numbers. \textcolor{black}{More specifically, the mass jump tends to happen for lower (higher) disc masses with longer (shorter) disc lifetimes to satisfy $t_{\rm PF}\sim t_{\text{acc},0}$.}
Moreover, the mass difference between low- and high-mass planets for a higher $\lambda$ case is smaller than that for a lower $\lambda$ case.  

Since the radial dependence of $\dot{M}_{\rm acc}$ is steeper for a lower $\lambda$ disc, and since $\dot{M}_F$ is assumed to be proportional to $\dot{M}_{\rm acc}$, 
we may naively expect a steeper $r$ dependence in planetary masses as well. 
However, the final core mass is set by the PIM that is independent of $\lambda$.
Therefore, if there is a sufficient pebble mass flux for a long enough period of time, it is possible that we don't find a strong dependence of planet formation outcomes on $\lambda$.
Our simulations, however, show that this is not the case and we indeed see the effects of $\lambda$ on planetary masses. 
We will discuss the dependence of planetary mass diversity on the magnetic lever arm $\lambda$ in the next subsection.

%
% \begin{figure*}[ht]
% \centering
% \includegraphics[width=\textwidth]{fig/meta_0_lambda_17.jpg}
% \caption{Similar to Fig.\ref{fig:meta_0_lambda_3} but with $\lambda = 17$.}
% \label{fig:meta_0_lambda_17}
% \end{figure*}
%
%At a glance, one can observe that wind-driven discs sustain significantly more diverse planets across almost all disc mass and $\alpha_{\text{total}}$ combinations. 
%However, the observed mass gradient in $\lambda = 3$ WD motivated us to continue investigating the dependence of planetary mass diversity on the magnetic lever arm.

\subsection{How \textcolor{black}{ $\lambda$ values affect the mass diversity in planetary systems}}
\label{subsec:lambda_mass}

\begin{figure*}[ht]
\centering
\includegraphics[width=\textwidth]{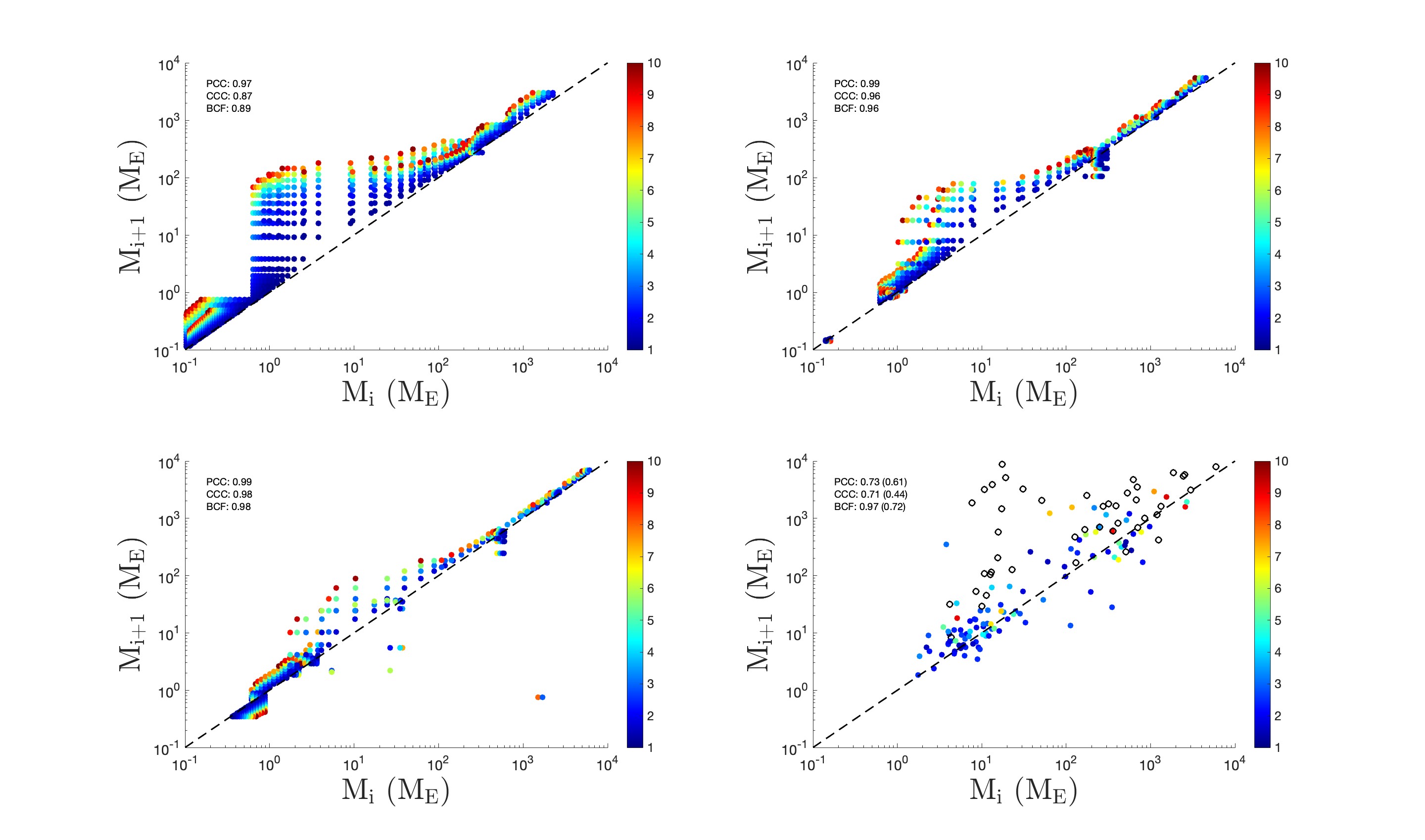}
\caption{Figure of \(M_{p,\text{out}}\) against \(M_{p,\text{in}}\) formed in wind-driven discs with ${\rm [Fe/H]} = 0$ and different \(\lambda\) values. 
\textcolor{black}{Top left: \(\lambda = 1.6\). Top right: \(\lambda = 3\). Lower left: \(\lambda = 17\). Lower right: observation.}}
\label{fig:PIP}
\end{figure*}

In this subsection, we will explore the effects of $\lambda$ on the similarity or diversity of mass distributions. 
Figure~\ref{fig:PIP} depicts the mass relationship between pairs of neighbouring planets for ${\rm [Fe/H]} = 0$, with the outer planet's mass plotted against that of the inner planet. 
The top left panel illustrates the scenario for \(\lambda = 1.6\), the top right panel for \(\lambda = 3\), the bottom left for \(\lambda = 17\), 
and the bottom right panel displays the observed planetary systems around single FGK stars with stellar masses ranging from 0.45 to 1.4 solar masses. 
We have selected the data from the NASA Exoplanet Archive from September 2024 for planets with mass uncertainties below 30\% (i.e., $\sigma_M/M<0.3$).
The colour of each marker represents the orbital period ratio between the outer and inner planets, 
while the unfilled circles in the bottom right panel are the observed pairs with ratios greater than 10. 

To generate this figure, we have grown 40 non-interacting protoplanetary cores per disc that are initially logarithmically equally spaced between 1 au and 100 au.  %1 AU to 100 AU within the same disc. 
Pairs of neighbouring planets with final orbital radii greater than 0.01 au and less than 10 au were then selected. 
%For each disc, we selected planetary pairs 5000 times. 
We have excluded pairs with the separation \(K_m<2\), where $K_m=\frac{a_{i+1}-a_i}{R_m}$ is defined 
%$K_m=\frac{\left| a_1 - a_2 \right|}{R_m}$ is defined 
as the orbital separation of two planets in terms of the mutual Hill radii ($R_m= \left(\frac{M_i+M_{i+1}}{3M_*}\right)^{1/3}\left(\frac{a_i+a_{i+1}}{2}\right)$).
%\(K = \frac{\left| a_1 - a_2 \right|}{R_m H}\).
The critical $K_m$ value is motivated by the observed planet pairs, for which $K_m$ ranges from 2.32 to about 60 and peaks around $10-15$.
%
%This exclusion was dictated by the assumption of the single-planet formation model, which allows that multiple planets can have closely spaced orbits around a star. 
Furthermore, only planet pairs with an orbital period ratio \(T_{i+1}/T_i \leq 10\) were considered\footnote{
Although an often adopted value is  \(T_{i+1}/T_i \lesssim 4\) \citep{Weiss2018,Mishra2021}, we have chosen a wider value because 
giant planets are often more widely separated \citep[$T_{i+1}/T_i\sim5-30$,][]{Rosenthal2024}.}.

By eye, we can see that $\lambda=17$ discs produce similar-mass neighbouring planets within this period ratio range, while smaller $\lambda$ discs such as $\lambda=1.6$ produce a larger fraction of neighbouring pairs with vastly different masses. 
More specifically, the low $\lambda$ discs form neighbouring planets with upto $\sim2$ orders of magnitude mass difference not only when planets are widely separated \(T_{i+1}/T_i \sim 10\) but also when planets are closely separated \(T_{i+1}/T_i \sim 2\).
On the other hand, although observed masses of neighbouring planets appear to be reasonably well-correlated, especially for lower-mass, closer pairs, there are also observed neighbouring planets with very different masses.   
%some work focuses on a more widely separated planets \(T_{i+1}/T_i \gtrsim 5.6\) \citep{Rosenthal2024}.
%Moreover, the distributions don't change dramatically by including planet pairs with \(T_{i+1}/T_i > 4\).}.
%to facilitate a better comparison with observational data. 
%
% The masses of these pairs were graphed, with the outer planet (indicated by \(i+1\)) on the y-axis and the inner planet (indicated by \(i\)) on the x-axis. 
% The colour of each marker represents the orbital period ratio between the outer and inner planets, the unfilled circles are the observation with ratios greater than 10. 
% For each panel, the Pearson Correlation Coefficient (PCC), Concordance Correlation Coefficient (CCC), and the Bias correction factor(BCF), 
% which is the ratio of PCC to CCC, were computed and presented. 
% For WDs across all \(\lambda\) values, the PCCs are approximately 1, which implies that almost all WDs produce planetary systems where \(M_{i+1} > M_i\). 
% However, for WDs with \(\lambda = 1.6\), 3, and 17, the CCCs are 0.88, 0.96, and 0.98, respectively. 
% The relatively low CCC for the \(\lambda = 1.6\) disc indicates a strong deviation from the diagonal line. 
% A correspondingly lower BCF, which further confirms the deviation, arises from the actual planet formation process.

To statistically check for the peas-in-a-pod trend, many studies calculate the Pearson Correlation Coefficient (PCC).  
Although the PCC measures the strength of the linear relationship between two variables, it is not sensitive to the deviation from the 45-degree line, such as shifts in the distribution of points above and below the line or differences in slope.  
To quantify these effects, we have also calculated the Concordance Correlation Coefficient (CCC), which is the product of the PCC and the Bias Correlation Factor (BCF). 
The latter represents how closely the data points cluster around the 45-degree line.
%For each panel, the Pearson Correlation Coefficient (PCC), Concordance Correlation Coefficient (CCC), and the Bias correction factor (BCF), 
%which is the ratio of PCC to CCC, were computed and presented.
These values are also noted in each panel.
For the lower-right panel, these values are shown for observed planet pairs with the period ratio \(T_{i+1}/T_i \leq 10\), 
while the corresponding values for all observed planet pairs are shown in parentheses. 

Observational data reveal a correlation between neighbouring planetary masses, with the PCC value of 0.73 and the corresponding p-value of $p<10^{-10}$.
%\textcolor{black}{***}\textcolor{black}{$4.9\times10^{-11}$ for filtered data and $2.1\times10^{-15}$ for all data. mass limit = 1e4}.
\cite{Otegi2022} studied peas-in-a-pod trends in radii, masses, densities, and period ratios for planets with $<120M_{\oplus}$ 
and found a clear correlation for mass with the PCC value of $\sim0.5$ and the p-value of $6\times10^{-4}$. 
%which are comparable to our values.
\cite{Mamonova2024} performed a similar study to \cite{Otegi2022}, and found no mass correlation in their main sample, 
but found a weak correlation of 0.35 by excluding giant planets with masses above $100M_{\oplus}$.
\cite{Rosenthal2024} found 0.476 for the mass correlation of giant planets. 
%
%is slightly larger than 
%0.476 obtained for giant planets \citep{Rosenthal2024} and 
%0.35 obtained by excluding giant planets with masses above $100M_{\oplus}$ \citep{Mamonova2024}. %  
%For planets with masses above $30M_{\oplus}$, 
%we have obtained the PCC value of 0.58, while for those with masses below $30M_{\oplus}$, the PCC value was 0.57.   
We have not found any increase in PCC values by excluding giant planets; the PCC value is 0.62 for $<30M_{\oplus}$ and 0.76 for $<100M_{\oplus}$. 
%\textcolor{black}{Correct} \textcolor{black}{Yunpeng, please confirm PCCs for 100ME.} 
The decreased PCC value for $<30M_{\oplus}$ compared to the full sample appears to be due to the lack of a group of planet pairs where one planet has $>30M_{\oplus}$ and the other has $<30M_{\oplus}$. 
%Since both of these values are lower than the PCC value for all, the observed trend shown here appears to disagree with what was suggested by \cite{Otegi2022,Mamonova2024}; the mass similarity trend is stronger for non-giant planets and becomes weaker by including giant planets.
%confirms that the mass similarity trend is stronger for non-giant planets and becomes weaker by including giant planets, as suggested by \cite{Otegi2022,Mamonova2024}.   
%Also, $68\%$ of planet pairs are above the diagonal line so that the outer planet is more massive than the inner one, while $32\%$ of pairs show the opposite trend.  
%planetary radii of Kepler planets (0.65)

On the other hand, the obtained CCC value of 0.71 for observed data indicates a poor correlation and a strong deviation from the diagonal line and 
thus a significant deviation from the peas-in-a-pod trend in neighbouring masses. 
The CCC values are 0.57 for $<30M_{\oplus}$ and 0.73 for $<100M_{\oplus}$, 
%\textcolor{black}{Yunpeng, please enter the value.} \textcolor{black}{Correct}
and thus again, our sample does not show a tighter correlation by excluding giant planets. 
%which indicates a poor but slightly stronger correlation compared to the all-mass case. 
%\textcolor{black}{***}\textcolor{black}{0.61} for $<30M_{\oplus}$, which indicates a \textcolor{black}{poor? strong?}
%\textcolor{black}{poor but stronger than all mass range CCC} correlation.
%This agrees with a visual impression that neighbouring masses appear to be more tightly correlated for closely-separated, low-mass planets. 

For WD discs with all \(\lambda\) values, the PCCs are approximately 1 and corresponding p values are $p\sim10^{-50}-10^{-100}$, 
%\textcolor{black}{$p<10^{-5}$????}\textcolor{black}{extremely low value(<1e-100)}, 
which implies that all WD models studied here show significant correlation in the mass of planet pairs.  
%However, the PCCs are not sensitive to the deviation from the daiagonal line such as the distribution of points above/below the 45-degree line 
%or the slope difference.  
%To quantify these effects, we have also calculated the CCCs, 
On the other hands, the CCCs are 0.88, 0.96, and 0.98 for \(\lambda = 1.6\), 3, and 17, respectively.
%\(M_{i+1} > M_i\). 
%However, the CCCs for WDs with \(\lambda = 1.6\), 3, and 17 are 0.88, 0.96, and 0.98, respectively. 
The relatively low CCC ($<0.90$) for the \(\lambda = 1.6\) disc indicates a poor correlation and a deviation from a similar-mass trend of neighbouring planets. 
%and a strong deviation from the diagonal line and 
%thus a significant deviation from the peas-in-a-pod trend in masses. 
%The figure indicates that the low $\lambda$ discs may form neighbouring planets with $\sim2$ orders of magnitude mass difference not only when planets are widely separated \(T_{i+1}/T_i \sim 10\) but also when planets are closely separated \(T_{i+1}/T_i \sim 2\).
%
For simulated planets, the mass correlations do not improve by excluding giant planets, because all of our simulations show very tight mass correlations for giant planets as seen in Figure~\ref{fig:PIP}.
Both PCCs and CCCs in fact decrease by excluding giant planets.
%We do not recover the observed trend that the mass correlation becomes weaker by including giant planets, because all of our simulations show very tight mass correlations for giant planets.
%Furthermore, all of our simulations show very tight mass correlations for giant planets.  
%Thus, by excluding giant planets, both PCCs and CCCs in fact decrease. %, which is the opposite of the observed trend. 
%This will be partly resolved by taking account of the mutual interactions, but such investigation is beyond the scope of this paper.  

Furthermore, observed data with \(T_{i+1}/T_i \leq 10\) show that $59\%$ of planet pairs are above the diagonal line so that the outer planet is more massive than the inner one, while $41\%$ of pairs show the opposite trend. 
%\textcolor{black}{Yunpeng, please confirm these numbers. I took them from the filtered data from your 23 Sep email by assuming that all data includes data with errors >30\%.} \textcolor{black}{Actually, data filtering based on error was applied to all the observation data. For observation data including T ratio $> 10$, the numbers are 0.68 and 0.32, while 0.59 and 0.41 are the numbers for pairs with T ratio $\leq 10$.}
As seen in Figure~\ref{fig:PIP}, for all of the WD cases, the outer planets tend to be more massive than the inner ones.  
The fractions of planet pairs above the diagonal line are $99\%$, $64\%$, and $65\%$ for \(\lambda = 1.6\), 3, and 17, respectively.
The fractions for $\lambda=3$ and $17$ are reasonably close to that estimated for the observed systems, 
though the observed planet pairs appear to be more symmetrically distributed around the diagonal line.  
This lack of symmetry is largely due to the lack of planet-planet interactions in our simulations.  
For interacting planetary cores, the mass distributions become more symmetric around the diagonal line as seen in, for example, 
Figure~7 of \cite{Mishra2021}.
%
% Furthermore, all of our simulations show very tight mass correlations for giant planets.  
% Thus, by excluding giant planets, CCCs in fact decrease, which is the opposite of the observed trend. 
% This will be partly resolved by taking account of the mutual interactions, 
% but the further investigation on this point is beyond the scope of this paper.  

We have also tested how PCCs and CCCs change depending on the maximum \(T_{i+1}/T_i\). 
For observed data, we find that CCCs never become $\geq0.90$ even when we change the maximum $T_{i+1}/T_i\leq$ so that neighbouring planetary masses are uncorrelated. 
%become $\geq0.90$ for $T_{i+1}/T_i\leq ***$, 
For simulated planets, even for $\lambda=1.6$, CCCs become $\geq0.90$ for $T_{i+1}/T_i\leq6$, so neighbouring planetary masses are generally well-correlated except for relatively \textcolor{black}{widely separated} pairs. 
%only closely-separated pairs are well-correlated.
Furthermore, by testing other $\lambda$ values, we have found that the CCCs become $>0.90$ for $\lambda\sim1.8$.
%Although we only show three cases of $\lambda$ here, we have tested 
%other $\lambda$ values as well and found that the CCCs become $>0.90$ for $\lambda\sim1.8$.
%\textcolor{black}{$\lambda\sim2.3$ Please confirm or modify this number}.
%\textcolor{black}{$\lambda\sim1.8$}.
%We have also tested how PCCs and CCCs change depending on the maximum \(T_{i+1}/T_i\). Even for $\lambda=1.6$, CCCs become $\geq0.90$ for $T_{i+1}/T_i\leq6$.
%Therefore, a statistically significant mass diversity is observed only for low $\lambda$ and for relatively widely-separated planet pairs.  
%Therefore, although we find that high $\lambda$ discs tend to produce similar-mass planet systems while low $\lambda$ discs can form more diverse-mass systems, a statistically significant mass diversity is observed only for low $\lambda$ and for relatively widely-separated planet pairs.  

In summary, although we find that high $\lambda$ discs tend to produce similar-mass planet systems while low $\lambda$ discs can form more diverse-mass systems, a statistically significant mass diversity is observed only for low $\lambda\lesssim 2$ and for relatively widely-separated planet pairs.  

Finally, for lower and higher metallicity cases, the general trends of these distributions stay the same; 
the deviation from the diagonal line becomes stronger for smaller $\lambda$. 
However, for the lowest metallicity of ${\rm [Fe/H]}=-0.5$, planet formation is less efficient and 
intermediate-mass planets with masses $10-100\,M_{\oplus}$ are scarce.  
For ${\rm [Fe/H]}=-0.5$, none of $\lambda$ cases lead to formation of diverse-mass planetary systems.  
%
%A correspondingly lower BCF, which further confirms the deviation, arises from the actual planet formation process.
%
% Observational data reveals a scattered distribution of planetary masses within exoplanetary systems. 
% However, a trend emerges from this diversity, indicated by a positive PCC, that typically, the mass of an outer planet exceeds that of its adjacent inner counterpart. 
% This observation aligns with the findings from our simulations. 
%Notably, simulations for protoplanetary discs with a \(\lambda = 1.6\) parameter generate closely spaced planetary pairs with mass differences spanning two orders of magnitude. 
%This significant mass disparity is primarily the consequence of pronounced radial variations in the local accretion rate within the disc. 
%Such variations suppress pebble accretion onto inner protoplanets with respect to outer ones. 
%the formation process of inner planets while accelerating the formation of outer planets.

\subsection{The occurrence rate of CJs in SE systems}
\label{subsec:SECJ}

\begin{figure}[ht]
    \centering
    \includegraphics[width=\columnwidth]{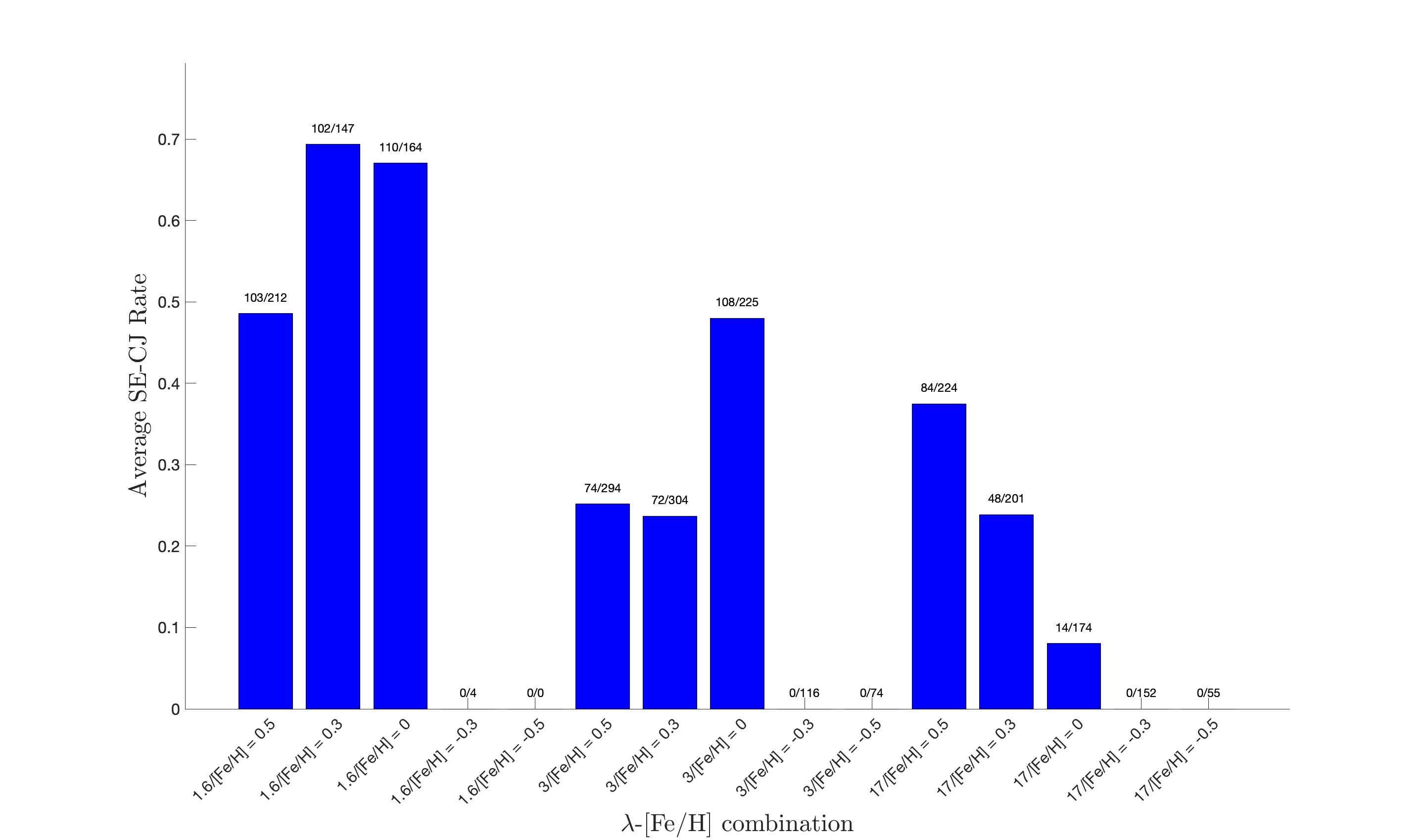}
    \caption{\textcolor{black}{Occurrence} rates of CJs in SE systems for various $\lambda$-${\rm [Fe/H]}$ combinations. 
    The fractions of SE-CJ pairs out of SE systems with outer companions are shown above the histograms as well. 
    The SE-CJ systems prefer high-metallicity discs (${\rm [Fe/H]}\geq0$) and/or low $\lambda$ discs.}
    \label{fig:SECJ}
\end{figure}

\cite{Bryan2024} calculated the frequency of outer giant planets in SE systems, and 
found that the frequency is $28.0^{+4.9}_{-4.6}\%$ for metal-rich stars (${\rm [Fe/H]}>0$) and 
$4.5^{+2.6}_{-1.9}\%$ for metal-poor stars (${\rm [Fe/H]}\leq0$).
Here, following \cite{Bryan2024}, we have estimated the occurrence rate of SE-CJ systems by calculating the frequencies of 
CJs with masses of $0.5-20\,M_J$ and orbital radii of $1-10\,$au 
in systems with a SE with mass of $1-20\,M_{\oplus}$ and orbital radius of $<1\,$au.

Figure~\ref{fig:SECJ} shows the occurrence rates of CJs in SE systems for various $\lambda$-${\rm [Fe/H]}$ combinations.
Since we have only focused on partiular disc masses and disc dissipation rates (represented by $\alpha_{\rm total}$), 
the actual occurrence rates of SE-CJ systems could be overestimated or underestimated per $\lambda$-${\rm [Fe/H]}$ combination.   However, \textcolor{black}{we still observed two general trends.} 
First, there is a clear preference of SE systems that are accompanied by CJs in metal-rich discs (${\rm [Fe/H]}\geq0$).
This is because only massive and/or long-lived discs tend to generate giant planets in metal-poor discs, 
and thus it is difficult to have SE-CJ systems.  
Second, there is also a clear preference of SE-CJ systems in low $\lambda$ discs; specifically, the average 
occurrence rates for metal-rich discs (${\rm [Fe/H]}\geq0$) are $62\%$, $32\%$, and $23\%$, for $\lambda=1.6$, $3$, and $17$, respectively.
\textcolor{black}{The high occurrence rate of SE-CJ systems in $\lambda = 1.6$ discs is caused by cores initially located beyond several au, which can reach the PIM fast enough in a high-metallicity environment to form CJs, whereas these cores did not grow into CJs in lower-metallicity environments.}
Since the occurrence rates of CJs in SE systems for $\lambda =1.6$ and 3 are higher than the estimated value by \cite{Bryan2024}, 
metal-rich discs with $\lambda \lesssim 3$ are expected to be good environments to generate SE-CJ systems.  
%
%In summary, a strong deviation from the diagonal line occurs for $\lambda\lesssim2$ from CCC values, 
%while high occurrence rates of SE-CJ systems in metal-rich discs occur for $\lambda\lesssim3$.  
%Thus, our simulations indicate that similar-mass and diverse-mass planetary systems are approximately separated at $\lambda\sim 2-3$. 
%Combining these results with those from the last subsection, our simulations indicate that similar-mass and diverse-mass planetary systems are approximately separated at $\lambda\sim 2-3$. 

%%%% REWRITE THIS PART ONCE A NEW FIGURE IS GENERATED BY REMOVING MULTIPLE COUNTINGS.
%%%% Need to discuss how lambda=1.6 leads to the largest difference in mass.   Also need to discuss the fraction of pls above and below y=x.

\subsection{How \textcolor{black}{ $\lambda$ values affect planetary migration}}
\label{subsec:migration}

\begin{figure}[h]
\centering
\includegraphics[width=\columnwidth]{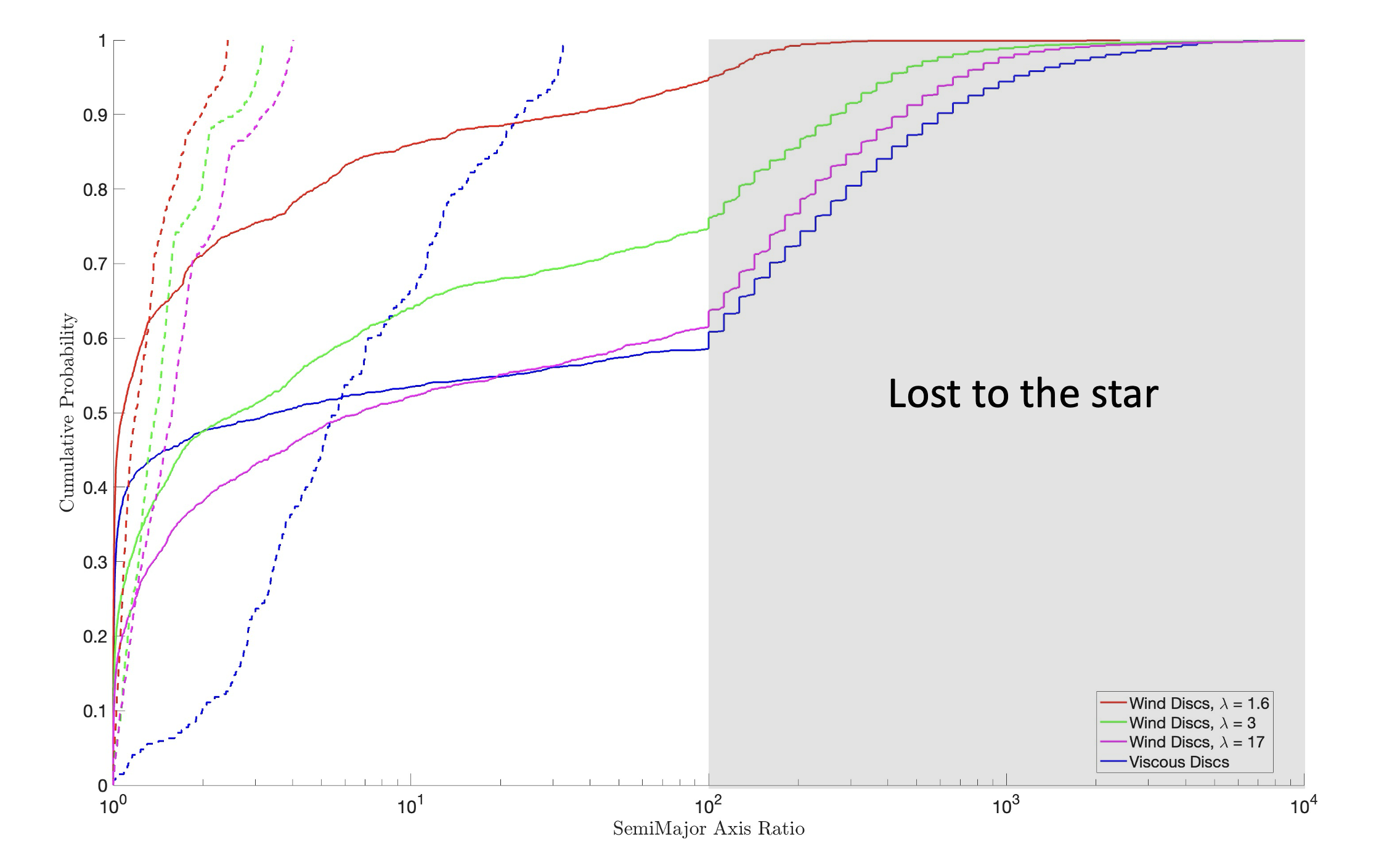}
\caption{Figure showing the ratios of the initial location to the critical location (\(r_{crit}\)) where giant planets started rapid gas accretion (solid lines), representing Type 1 migration. Dashed lines show the ratios of \(r_{crit}\) to the final location, representing Type 2 migration. Red, green, and magenta lines represent planets formed in wind-driven discs with \(\lambda = 1.6\), 3, and 17, respectively, while blue lines represent those formed in purely viscously evolving discs.
 }
\label{fig:type1+type2}
\end{figure}

Here, we will discuss how \(\lambda\) values affect planet migration.
%Here, we will further analyse how \(\lambda\) values affect planet migration.
%the migration of giant planets. 
%Figure~\ref{fig:type1+type2} shows planet migration efficiencies for both type I (solid curves) and type II (dashed curves) migrations. 
Figure~\ref{fig:type1+type2} shows the extent of planet migration by semimajor axis ratios 
for both type I (solid curves) and type II (dashed curves) migrations. 

For type I migration, we have plotted the migration extent \textcolor{black}{for cores that} never became giant planets and for those becoming giants. 
For the former, the ratios of the initial to the final semimajor axes are used, 
while for the latter, the ratios of their initial location to the location $r_{\rm crit}$ 
where they started rapid gas accretion (i.e., where $M_{\text{core}} = M_{\text{env}}$) are chosen. \textcolor{black}{If a core is lost to the star, its final location is assumed to be at the disc's inner edge, $0.01\,$au, which results in step-like increments in the shaded region.} %, marked as \( r_{crit} \),
The median semimajor axis ratios are approximately $7$, $2.5$, and $1.2$ for $\lambda=17$, $3$, and $1.6$, respectively. 
Thus, the figure shows that type I migration is less efficient in a lower $\lambda$ disc. 
%that has a flatter surface mass density profile.
This is partly because planet migration itself slows down in a flat surface mass density disc \citep{Ogihara2018}, 
and partly because planet formation is slower in such a disc due to the lower mass flux.     

These points can be seen in two extremes of the distributions as well.  
On one end, the shaded region represents the cores lost to the central star.
%shows the step-like increments, which represent the cores lost to the central star.
The fractions of lost planets are approximately $5\%$, $25\%$, and $40\%$, for $\lambda=1.6$, $3$, and $17$, respectively.
So fewer cores were lost in low $\lambda$ discs compared to high $\lambda$ ones. 
%due to the combined effects of less efficient planet formation 
%and a flatter surface mass density. 
%
On the other end, a large fraction of planets with the semimajor axis ratio near 1 arises from planets that barely grew.  
Since the fraction is the largest for $\lambda=1.6$ and the smallest for $\lambda=17$, 
the figure further confirms that planet formation is slower in the lower-$\lambda$ discs.

Interestingly, despite that the planet loss rate is high ($\sim40\%$) and comparable to that of the $\lambda=17$ case, 
VE discs also have a large fraction of non-migrating planets ($\sim 30\% $), which is higher than the values for $\lambda=17$ and $\lambda=3$ 
($\sim10\%$ and $\sim20\%$, respectively) though is not as high as $\lambda=1.6$ ($\sim40\%$). 
This is because planet formation is in fact slow in VE discs because the pebble scale height is larger compared to WD discs 
due to the higher disc's viscosity. % assumed in the same disc.  % 
The large fraction of non-migrating planets arises mostly from the outer disc where formation timescales are long. 
On the other hand, VE discs also lose many planetary cores to the central stars because the PIM is higher due to the higher viscosity\textcolor{black}{,} 
and thus cores migrate further \textcolor{black}{towards} the star before reaching the PIM, accreting gas, and switching to slower type II migration. 
As it can be seen from Equation~\ref{eq: mig2}, since type I migration becomes faster as the mass increases while type II becomes slower, planet migration slows down once the gas accretion starts.  
%
%furhter indicating that more cores survived type 1 migration due to the increasingly flatter surface mass density in the inner region. 
%
%
%In figure~\ref{fig:type1+type2}, all the surviving giant planets were selected, and the ratios of their initial location to the location 
%where they started to rapidly accrete gas, marked as \( r_{crit} \), were plotted in solid lines. 
%These ratios represent the extent of protoplanet migration during the Type 1 migration stage. The ratios of \( r_{crit} \) to the final location, 
%plotted in dashed lines, were then calculated to measure the Type 2 migration of giant planets.

For type II migration, the semimajor axis ratio in Figure~\ref{fig:type1+type2} represents the ratio of the critical radius for rapid gas accretion $r_{\rm crit}$ to the final location.
Type II migration is also slower in WD discs compared to VE discs because \(\alpha_{\text{SS}}\) is smaller in WD discs. 
In other words, giant planets formed in VE discs migrate significantly after/during the gas accretion phase, 
while those formed in WD discs tend not to migrate much during this phase (typically less than a factor of two in terms of the semimajor axis ratio).   
Type II migration is also slower in lower $\lambda$ discs compared to higher $\lambda$ ones, partly because a flatter surface mass density profile leads to a smaller torque as in type I migration, 
and partly because giant planet formation takes longer in these discs\textcolor{black}{,} and thus \textcolor{black}{a} smaller amount of disc masses remain by that time. 
% The step-like increments imply that the cores were lost to the star. For a lower lambda disc, fewer cores were lost, 
% indicating that more cores survived Type 1 migration due to the increasingly flatter surface mass density in the inner region. 
% Type 2 migration was also slower in WD compared to VD because the \(\alpha_{\text{SS}}\) was smaller in WD. 
% The result is consistent with \citep{Ogihara2018}. VDs naturally have the steepest and highest \(\alpha_{\text{ss}}\), resulting in the fewest planets surviving in them.

\subsection{How \textcolor{black}{ $\lambda$ values affect the overall planetary mass}}

\begin{figure}[h]
\centering
\includegraphics[width=\columnwidth]{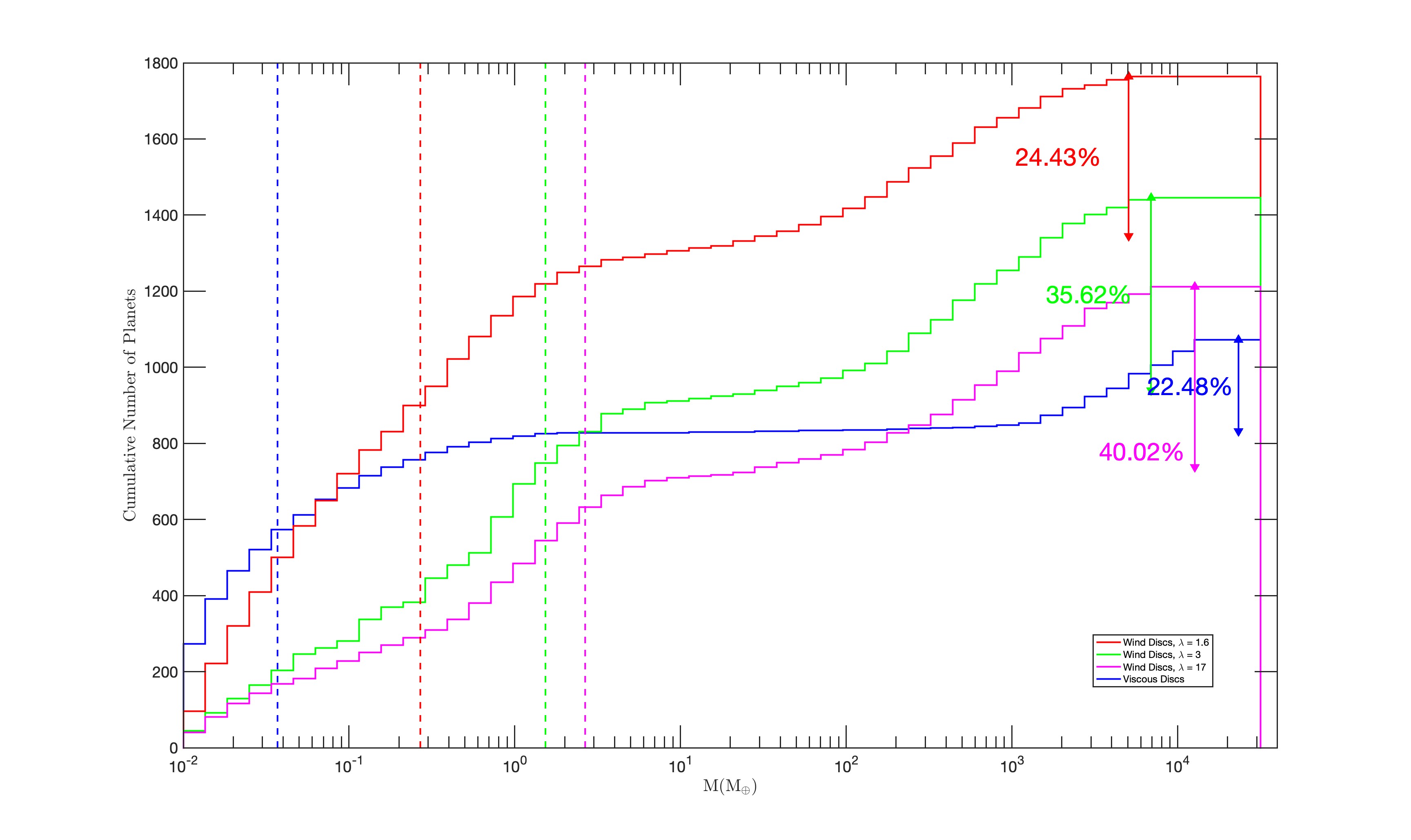}
\caption{\textcolor{black}{Cumulative} number of planets formed in wind-driven discs with \(\lambda = 1.6\), 3, and 17, shown in red, green, and magenta lines, respectively. The blue line represents the planets formed in purely viscously evolving discs. The arrow lines and numbers represent the percentage of planets with a final mass greater than 30 \textcolor{black}{\(M_{\oplus}\)}, and the vertical dashed lines are the median values of planetary mass for each type of disc.}
\label{fig:M cumulative}
\end{figure}

To show how \(\lambda\) values affect the overall final planetary mass distribution, we have plotted the cumulative mass distributions of 
surviving planets in Figure~\ref{fig:M cumulative}. 
%we show the mass of surviving planets and the cumulative number of planets in Figure~\ref{fig:M cumulative}. 
%For each coloured line, 1800 single core formations are simulated (9 discs \(\times\) 40 cores \(\times\) 5 metallicities). 
Each coloured line includes 1800 single core formation simulations (9 discs \(\times\) 40 cores \(\times\) 5 metallicities). 
The arrows and numbers represent the percentage of surviving giant planets with a final mass greater than 30 \(M_{\oplus}\), 
while the vertical dashed lines are the median values of planetary mass for each type of disc. 
%The red, green, and magenta lines represent WD discs with \(\lambda = 1.6, 3\), and \(17\), respectively. 
The WD discs with \(\lambda = 1.6, 3\), and \(17\) are shown in red, green, and magenta, respectively, while 
the VE discs are shown in blue.

The figure indicates that a larger number of planets survived in lower $\lambda$ discs and 
that surviving planetary masses tend to be lower in such discs.  
These two outcomes are intertwined with each other.  
As described earlier, the discs with lower \(\lambda\) values are more efficient at preserving planets 
due to slower planet formation and migration, but this also means that the planetary masses tend to stay lower.  
For $\lambda=1.6$, roughly $67\%$ of surviving planets have $<M_{\oplus}$ and the median mass is $\sim0.3M_{\oplus}$, 
while for $\lambda=17$, $\sim42\%$ of surviving planets have $<M_{\oplus}$ and the median mass is $\sim3M_{\oplus}$.
%Since planet formation is slower, the lower $\lambda$ values also result in fewer giant planets per disc, which can be seen 
%from the percentage of surviving giant planets. 
%Despite that most planets survived, the fraction of giant planets in $\lambda=1.6$ discs is $\sim24.4\%$.
The difference in the numbers of surviving planets is in fact mainly caused by low-mass planets and 
the numbers of surviving giant planets are comparable to one another among WD discs ($\sim400-500$ out of 1800 cores).
%For the $lmabda=1.6$ case, roughly $67\%$ of surviving planets have $<M_{\oplus}$ and the median mass is $<0.3M_{\oplus}$.
%Since planet formation is slower and migration is less efficient in low $\lambda$ discs, 
%roughly $67\%$ of surviving planets for the $\lambda=1.6$ case have $<M_{\oplus}$.
%The lower $\lambda$ discs also have a lower median mass of planets as indicated by vertical dashed lines.  
%The lower local accretion rate also slows down the formation stage, resulting in these discs forming the fewest giant planets and 
%having the lowest median mass among all wind discs. 
%Conversely, discs with higher \(\lambda\) values lost many protoplanets (approximately $33\%$ for $\lambda=17$) 
%due to more efficient formation and migration.   
%Nevertheless, higher $\lambda$ discs have higher median planetary masses and higher percentages of surviving giant planets. 
%
%The figure also indicates that a higher $\lambda$ (i.e., the faster formation combined with the steeper surface mass density radial profile) 
%leads to severe loss of planets.
%However, the faster formation combined with the steeper surface mass density radial profile also leads to severe loss of planets, 
%While nearly all planets survived in \(\lambda = 1.6\) discs, approximately 33\% of the cores are lost in \(\lambda = 17\) discs. 

Finally, although VE discs form the heaviest planets, the survival rate of giant planets and the median planetary mass 
are the lowest among all discs, owing to efficient migration and significant planet loss. 
%the giant planet formation rate and median planetary mass are the lowest among all discs, 
%owing to intense planet loss due to efficient migration.
This is primarily because the disc turbulence in VE discs is scaled with $\alpha_{\rm total}$\textcolor{black}{,} which is higher than $\alpha_{\rm SS}=10^{-4}$ for WD discs. 
%In summary, the survival rate of giant planets in WD discs is higher than in VE discs by 1.95\%, 13.14\%, and 17.54\% 
%in \(\lambda = 1.6, 3,\) and \(17\) discs, respectively. 
%In summary, WD discs with lower \(\lambda\) values more efficiently preserve planets, partly because of less efficient planet formation 
%and partly because of less efficient planet migration.
%We will further discuss the effects of planet migration in the next subsection. 

\subsection{How \textcolor{black}{different metallicity values affect overall planetary mass and orbital radius}}\label{subsec:metallicity}

\begin{figure*}[h]
\centering
\includegraphics[width=\textwidth]{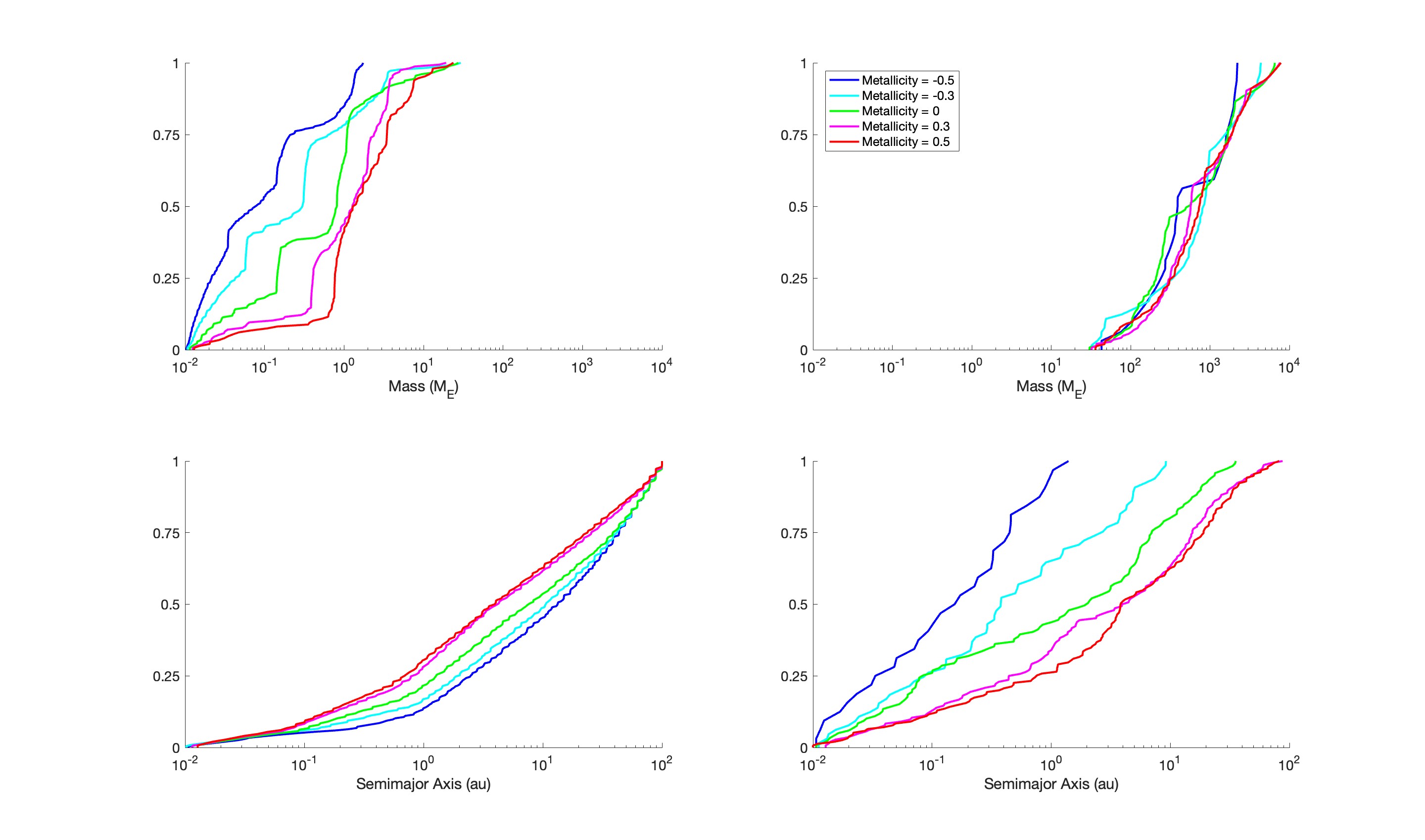}
\caption{\textcolor{black}{Effects} of metallicity on planetary masses and orbital radii for \(\lambda = 3\). 
The top-left and top-right panels show the cumulative distribution of planetary mass 
for low-mass planets (\(M < 30M_\oplus\)) and giant planets (\(M > 30M_\oplus\)), respectively. 
%The top-left and top-right panels show the cumulative distribution function (CDF) of planetary mass 
%for low-mass planets (\(M < 30M_\oplus\)) and giant planets (\(M > 30M_\oplus\)), respectively. 
The lower-left and lower-right panels display the corresponding cumulative distribution of their orbital radii.}
\label{fig:metallicity analysis}
\end{figure*}

Here, we explore how metallicity affects planetary masses and orbital radii.
Figure~\ref{fig:metallicity analysis} shows the cumulative distributions of planetary masses (top panels) and their orbital radii (bottom panels) 
for low-mass (\( M < 30M_\oplus \), left panels) and high-mass (\( M > 30M_\oplus \), right panels) planets.
%The top-left and top-right panels show the cumulative distribution function (CDF) of planetary mass for low-mass planets (\( M < 30M_\oplus \)) 
%and giant planets (\( M > 30M_\oplus \)), respectively. 
%The lower-left and lower-right panels show the corresponding CDFs of their orbital radii. 
These planets were formed in WD discs with \(\lambda = 3\).

In order to quantitatively test the metallicity dependence, we performed the KS test on all cumulative distribution function (CDF) pairs shown 
in Figure~\ref{fig:metallicity analysis}. 
The results are \textcolor{black}{summarised} in Figures~\ref{fig:ks_test_gt30} and \ref{fig:ks_test_lt30} for giant planets and low-mass planets, respectively. 
In each figure, the top-right triangle, composed of grids, represents the test results for orbital radius, 
while the bottom-left triangle represents the results for planetary mass. 
%The p-values and d-values are shown for each metallicity sample pair. 
The two-sample KS statistic $d$ and the corresponding probability $p$ are shown for each metallicity sample pair.  
Grids representing orbital radius with p-values less than 0.05 are highlighted in red, while those showing planetary masses are highlighted in cyan.

Figure~\ref{fig:metallicity analysis} shows that the mass distributions of low-mass planets (top left panel) vary 
depending on the metallicity, while those of high-mass planets (top right panel) are largely independent of the metallicity.
This is confirmed by the KS tests.  
%Regarding the planetary mass distribution, all CDF pairs for each metallicity combination rejected the null hypothesis for low-mass planets, 
For low-mass planets, the null hypothesis is rejected for all pairs of metallicities, 
suggesting that the disc metallicity significantly altered the mass distribution. 
Since the core accretion stage dominates the formation of such planets, higher metallicity leads to higher planetary mass.  
%However, this trend is not the same for giant planets. 
This trend, however, does not hold for giant planets.  
Although some metallicity combinations result in the p-value of $<0.05$, we cannot rule out the null hypothesis for the others.
This is understandable because the PIM is independent of the metallicity (see Equation~\ref{eq:pebble-isolation-mass}) 
so gas accretion occurred on similar-mass cores.
It should be noted that, although these curves appear similar, the number of planets included differs depending on metallicities.
For the lowest metallicity environment, giant planet formation occurred only in relatively massive and long-lived discs (discs 4, 7, and 8), 
while for the highest metallicity environment, it happened in all the discs except for short-lived discs (discs 2 and 3).
%The envelope mass dominates the mass of giant planets, and protoplanets formed 
%in low-metallicity (e.g., metallicity = -0.5) discs only trigger gas accretion in discs 4 and 7 (i.e., in massive and relatively long-lived discs). %disc 4 and disc 7. 
%Although the curves appear similar, the number of planets included differs.
A kink seen for certain mass ranges in the distribution (e.g., \(10^{2.2} \, \textcolor{black}{M_\oplus} \sim 10^{3} \, \textcolor{black}{M_\oplus}\)
) for \textcolor{black}{$[\text{Fe/H}]= -0.5$} discs) 
arises due to rapid gas accretion so that few planets within that range of masses exist. 
%Gas accretion rapidly increases planetary masses, so that few planets within that range of masses exist, %planets within that range of masses do not exist, 
%resulting in a 'flat' section for certain mass ranges in the distributions (e.g., \(10^{2.2} M_E \sim 10^{3} M_E\) for metallicity = -0.5 discs). 
%The presence of this 'flat' sections in the cumulative distributions at different mass ranges lead to the rejection of the null hypothesis for some metallicity 
%combinations (e.g., metallicity = -0.5 and metallicity = -0.3). 
%Other than that, once giant planets are formed, their mass distributions show no significant difference across different initial disc metallicities.

The bottom panels of Figure~\ref{fig:metallicity analysis} show that the orbital radius distributions for low-mass and high-mass planets 
exhibit an opposite trend.
For low-mass planets, higher metallicity environments lead to smaller orbital radii, while for high-mass planets, 
higher metallicity environments lead to larger orbital radii.  
Both of these are the results of more efficient planet formation in the higher metallicity environment. 

For high-mass planets, the higher-metallicity environment leads to faster formation of cores\textcolor{black}{,} and thus gas accretion tends to be triggered at larger orbital radii.
In fact, as seen from the bottom right panel, roughly $75\%$ of giant planets in the higher metallicity environments (0.3 and 0.5) have cold Jupiters with orbital radii $>1\,$au 
and nearly $40\%$ of giants survived beyond $10\,$au.  
The lower-metallicity environment, on the other hand, results in slower formation and thus more migration of cores before triggering gas accretion. 
More than $40\%$ of giant planets in the lowest-metallicity environment formed giants within $0.1\,$au.

For low-mass planets, on the other hand, the distributions are dominated by planets \textcolor{black}{that} were initially located in the outer part of the disc 
and thus were growing slowly.  
Indeed, more than $50\%$ of the low-mass planets orbit beyond $10\,$au in the lowest metallicity environment, while 
this number decreases to $30\%$ in the highest metallicity environment\footnote{\textcolor{black}{This apparent decrease in the fraction of cold low-mass planets for higher metallicity environments is caused by the fact that more planetary cores become giant planets rather than low-mass planets due to a higher pebble mass flux.}}.
%about $30\%$ of the low-mass planets orbit beyond $10\,$au in the highest metallicity environment. 

The KS tests show that we can reject the null hypothesis for most cases of the radius distributions. 
Thus, our model predicts a metallicity gradient in the orbital distribution of gas giants in the quiescent environment where 
the gravitational perturbation effects were weak and the orbital distribution of giant planets just before the disc dissipation is well-preserved.   

Finally, for lower and higher $\lambda$ values, the overall trends are similar.  
For high-mass planets, both mass and orbital radius distributions for each metallicity have weak dependences on $\lambda$, and thus the distributions look very similar to Figure~\ref{fig:metallicity analysis}.
For low-mass planets, masses tend to be lower and orbital radii tend to be larger for lower $\lambda$ discs.   
This trend reflects the fact that planet formation is slower and migration is less efficient in lower $\lambda$ discs.   

% On the other hand, the distribution of orbital radii for low-mass and giant planets exhibits an opposite trend.   %a reversed trend. 
% In the low-metallicity discs, more protoplanets initially located in the outer disc region couldn't grow significantly. 
%These low-mass planets contribute to the orbital radius distribution, resulting in more than 50\% of the low-mass planets orbiting beyond 10 au 
%in the lowest metallicity environment. 
% However, in the high-metallicity discs, those protoplanets were able to eventually form giant planets, 
% leading to only about 30\% of the low-mass planets orbiting beyond 10 AU. 
% Furthermore, significant differences due to metallicity are required to reject the null hypothesis, suggesting that the orbital radius distribution 
% for low-mass planets depends only weakly on disc metallicity.
%
% Finally, the distribution of orbital radii for giant planets rejected the null hypothesis for almost all metallicity combinations except the [0.3, 0.5] pair. 
% As stated above, giant planets are more easily formed in the outer region of the disc in higher metallicity environments. 
% These giant planets form later and therefore experience mild migration. Thus, higher metallicity discs tend to form giant planets at wider orbits.

\section{Discussion}
\label{sec:Discussion}
\textcolor{black}{In this section, we will discuss the effects of multiple-core formation and the \textcolor{black}{radially dependent} initial core masses.}
%the disc evolution model. 
%
\subsection{Effects of multiple-core planet formation}
%Our model has adopted multiple assumptions.  In this section, we discuss some of these effects.
Since the main purpose of our work is to investigate the effects of $\lambda$ on formation and migration of planets, we have adopted a single-core planet formation model. However, multiple-core effects and their associated influences are likely important for planet formation in realistic protoplanetary discs.  
%In this section, we discuss potential effects of multiple-core planet formation.  

%One such effect is a dynamical one.  
In our model, all the planets are implicitly assumed to have nearly circular and coplanar orbits. 
The dynamical interactions among (proto-)planets can change eccentricities and inclinations of orbits significantly, 
which could affect both formation and the final orbital distribution of planets.
For example, we adopt Equation~\ref{eq:pebble_accretion_efficiency} for the pebble accretion efficiency, but the formula does not take account of 
the effects of the randomness of orbits.  For multiple-core simulations, it would be more appropriate to adopt an efficiency formula such as \cite{Ormel2018b} 
that includes the effects of eccentricities, inclinations, and disc turbulence.  For zero eccentricity and inclination, these two efficiencies 
agree reasonably well with each other \citep{Matsumura2021}.

As the gas disc dissipates, the dynamical interactions among planets can lead to scattering, orbital crossing, and/or 
removal of some of the planets from the system, which drive eccentricities and inclinations of remaining planets 
\citep[e.g.,][]{Chatterjee2008,Juric2008,Matsumura2010a}. 
When the eccentricity of a planetary orbit is high enough so that the pericentre of the orbit is close to the central star, 
the tidal interactions between the star and the planet can circularise its orbit, leading to formation of hot or warm Jupiters.
Since higher metallicity discs tend to have a larger number of giant planets, such dynamical instabilities followed 
by the tidal circularisation of the orbit and the formation of a close-in planet may happen more frequently in a higher metallicity environment.
Therefore, the orbital distribution trend seen in Figure~\ref{fig:metallicity analysis} is likely to change 
when considering multiple-core planet formation.
Furthermore, scattering among planetary cores is likely to transform the one-sided neighbouring mass distribution seen in Figure~\ref{fig:PIP} 
into something more symmetric around the diagonal line, as seen in the observed systems.

%Independent of the dynamical effects, the planet formation outcomes presented in this work may change when we consider the growth of multiple cores. 
Independent of the dynamical effects, the planet formation outcomes may also change when we consider the growth of multiple cores. 
Once a protoplanetary core reaches the PIM, it is expected to perturb the disc enough to create a pressure bump in the upstream 
\textcolor{black}{that} could trap the incoming flux of pebbles \citep[e.g.,][]{Ormel2017,Johansen2017}.
Such an effect can deprive protoplanets interior to the PIM core of the pebble flux and potentially create the dichotomy in the 
planetary system \citep{Morbidelli2015}. 
Our simulations have not taken this effect into account.
However, in our simulations, there were no surviving cores %that increased the masses \textcolor{black}{by more than a factor of **} \textcolor{black}{no cores survive at all}
inside the first surviving planet that reached the PIM.  
Therefore, we don't expect a large difference in planet formation outcomes for inner protoplanets, even when we take this effect into account.
The PIM core may also influence the growth of outer protoplanets. 
A recent work by \cite{Lau2024} has shown that the second generation of planets can be formed in a pressure bump due to the pebble accumulation there. 

\subsection{Effects of initial core masses}
%To justify the choice of a fixed embryo mass of \(0.01 \, M_\oplus\) across the disc, 
\textcolor{black}{In this work, we have assumed constant initial core masses across the disc for simplicity.  
However, it is possible that initial core masses are radially dependent.
We have tested two types of radially dependent initial embryo masses: those determined by the streaming instability \citep{Liu2020} and those equal to the pebble accretion onset masses \citep{Ormel2017}.} 

\textcolor{black}{
For the former, the streaming instability simulations predict that the  most massive core masses are given by
\begin{equation}
    M_{\rm core,\,in} = 2 \times 10^{-3} \left( \frac{\gamma_{\rm SI}}{\pi^{-1}} \right)^{3/2} \left( \frac{\hat{h}_g}{0.05} \right)^3 \left( \frac{M_\star}{0.1 M_\odot} \right) M_\oplus.
    \label{eq:M_init_SI}
\end{equation}
Where \(\gamma_{\rm SI} = 4 \pi G \rho_g / \Omega_K^2\) is the relative strength between self-gravity and tidal shear, \(\rho_g = \Sigma_g / (\sqrt{2 \pi} \hat{h}_g r)\) is the gas volume density.
Since the core mass depends on the surface mass density and thus $\lambda$, the initial core mass varies not only radially but also for different disc models.  
A typical range of core masses \textcolor{black}{is} $\sim10^{-7}-10^{-4}\,M_{\oplus}$ at $\sim1\,{\rm au}$, and \textcolor{black}{it reaches} up to $\sim10^{-2}-10^{-1}\,M_{\oplus}$ at $\sim100\,{\rm au}$.
}

\textcolor{black}{
For the latter, the pebble accretion onset mass is given by 
\begin{equation}
M_{\rm core,\,in} = \tau_s\eta^3M_*\, .
\label{eq:M_init_onset}
\end{equation}
This core mass also varies both radially and for different disc models via $\eta$ and $\tau_s$.
A typical range of core masses is $\sim10^{-4}\,M_{\oplus}$ at $\sim1\,{\rm au}$, and \textcolor{black}{it reaches} up to $\sim10^{-1}\,M_{\oplus}$ at $\sim100\,{\rm au}$.
}

\textcolor{black}{
For both types of initial core masses, the main findings of this work remain largely unchanged.
For \(\lambda = 1.6\) with the SI-induced initial core masses, however, discs produce less SEs in the inner disc because planet formation is very slow due to very low embryo masses (down to $\sim10^{-7}\,M_{\oplus}$). Nevertheless, \(\lambda = 1.6\) discs still yield planetary systems with the greatest mass disparity among planets.
Another difference is that giant planets are formed more efficiently in the outer disc, beyond a few tens of au, since initial core masses tend to be higher there in both cases.
}

\section{Summary}
\label{sec:Summary}

%In this paper, we studied planet formation in wind-driven discs via pebble accretion. 
%The effects of various parameters on planet formation, such as initial disc mass, \(\alpha_{\text{total}}\), metallicity, and \(\lambda\), were analysed. 
%In this paper, we have investigated a hypothesis that the fast-wind (or efficient-wind) accretion discs lead to formation of close-in, 
%similar-mass planetary systems as seen in peas-in-a-pod systems, while the slow-wind discs lead to planetary systems with greater mass differences. 
%In this paper, we have investigated a hypothesis that the efficient WD accretion discs (i.e., high $\lambda$ discs) 
%lead to formation of close-in, similar-mass planetary systems as seen in peas-in-a-pod systems, 
%while the less efficient WD discs (i.e., low $\lambda$ discs) lead to planetary systems with greater mass differences. 
%Specifically, we have studied single-core planet formation in wind-driven accretion discs via pebble accretion and 
%explored the effect of the magnetic lever arm parameter $\lambda$ on the outcome of planet formation and migration.
%We adopted the pebble accretion model developed by \cite{Ida2016} and the wind-driven disc model by \cite{Tabone2022a}. 
%
In this paper, we have studied single-core planet formation in WD accretion discs via pebble accretion and 
explored the effect of the magnetic lever arm parameter $\lambda$ on the outcome of planet formation and migration.
We adopted the pebble accretion model developed by \cite{Ida2016} and a hybrid WD and VE disc model by \cite{Tabone2022a}. 
%Our main motivation was to investigate a hypothesis that the efficient WD accretion discs (i.e., high $\lambda$ discs) 
%lead to formation of close-in, similar-mass planetary systems as seen in peas-in-a-pod systems, 
%while the less efficient WD discs (i.e., low $\lambda$ discs) lead to planetary systems with greater mass differences. 

%Our main findings are summarised as follows.
%
A lower value of $\lambda$ corresponds to a disc with a higher mass ejection rate, \textcolor{black}{which} results in a steeper radial dependence of 
the disc mass accretion rate, a lower stellar mass accretion rate, and a flatter surface mass density profile \citep{Tabone2022a}.
% Overall, our simulations show that, for higher $\lambda$ discs, 1. planet formation is faster, 
% 2. both type I and type II migration is faster, and 3. neighbouring planetary masses tend to be similar to each other, 
% and vice versa for lower $\lambda$ discs. 
%
Overall, there are three main differences in formation and evolution of planets in different $\lambda$ discs. 
First, planet formation is slower in lower $\lambda$ discs compared to higher $\lambda$ ones.  
This is because the disc mass dissipates faster and the pebble mass flux becomes lower in a lower $\lambda$ disc.

Second, both type I and type II migrations are slower in a lower $\lambda$ disc 
(see Figure~\ref{fig:type1+type2} and Section~\ref{subsec:migration}). 
In such a disc, type I migration is slowed largely because of a flatter surface mass density profile\textcolor{black}{,} as proposed by \cite{Ogihara2018}. 
Moreover, since type I migration becomes faster with planetary mass (see Equation~\ref{eq: mig1}), 
less efficient planet growth in a lower $\lambda$ disc leads to less migration overall. 
By adopting a new migration formula by \cite{Kanagawa2018}, type II 
migration scales with type I migration\textcolor{black}{,} as \textcolor{black}{can be} seen in Equation~\ref{eq: mig2}, and thus 
a flatter surface mass density also leads to slower type II migration.  

Third, a higher value of $\lambda$ leads to a system of similar-mass planets, while a lower value of $\lambda$ leads to a greater variety of planetary masses across the disc (see Figure~\ref{fig:PIP}). 
\textcolor{black}{More specifically, although a planetary system with a mass jump can be found for all $\lambda$ cases as long as $t_{\rm PF}\sim t_{\text{acc},0}$ (see Section~\ref{subsec:growth_tracks}), the mass jump tends to be larger for lower $\lambda$ discs.}
We also found that the frequencies of SE-CJ systems are higher in lower $\lambda$ and/or higher metallicity discs (see Section~\ref{subsec:lambda_mass}).  
The latter agrees with the observed preference of such systems around metal-rich stars \citep{Bryan2024}. 
Combining the preference for similar-mass neighbouring planets and efficient migration in high $\lambda$ discs, it is expected that such discs tend to lead to \textcolor{black}{the} formation of close-in, peas-in-a-pod, multiple-planet systems. 
From the discussions in Sections~\ref{subsec:lambda_mass} and \ref{subsec:SECJ}, we find that similar-mass and diverse-mass planetary systems are approximately separated at $\lambda\sim2-3$.

Furthermore, our simulations also indicate the following effects of the stellar metallicity on formation and migration of planets. 
First, a higher-metallicity disc leads to a more massive and closer-in rocky planet compared to a lower-metallicity disc (see the left panels of Figure~\ref{fig:metallicity analysis}). 
This is because formation is faster\textcolor{black}{,} and thus a (proto-)planet has a longer time to migrate in the disc.
Second, \textcolor{black}{giant planet masses are expected to be} largely independent of the metallicity (see the upper right panel of Figure~\ref{fig:metallicity analysis}) if the pebble isolation mass is independent of the metallicity\textcolor{black}{,} as assumed here. \textcolor{black}{Third, if the dynamical evolution is quiescent, the metallicity gradient may be seen for orbital periods of giant planets.} We expect giant planets in lower-metallicity discs tend to have shorter orbital periods compared to those in higher-metallicity discs (see the lower right panel of Figure~\ref{fig:metallicity analysis}) \textcolor{black}{because the core formation takes longer in a lower metallicity disc\textcolor{black}{,} and thus cores migrate further before gas accretion}. 

\begin{acknowledgements}
We thank an anonymous referee for the prompt feedback. We thank Ramon Brasser, Man Hoi Lee, Tommy Chi Ho Lau, Shigeru Ida, and Aurora Sicilia-Aguilar for various discussions relevant to this work. SM would also like to thank the hospitality of the Konkoly Observatory, where part of this work was conducted. 
\end{acknowledgements}

\bibliographystyle{aa} 
\bibliography{pebble_accretion_references}

\begin{appendix}
\onecolumn 

\section{KS test results}
\label{sec:appendixA}
\textcolor{black}{The following two figures summarise the KS test results discussed in Section~\ref{subsec:metallicity}.}

\begin{figure}[h!]
\centering
\includegraphics[width=0.98\textwidth]{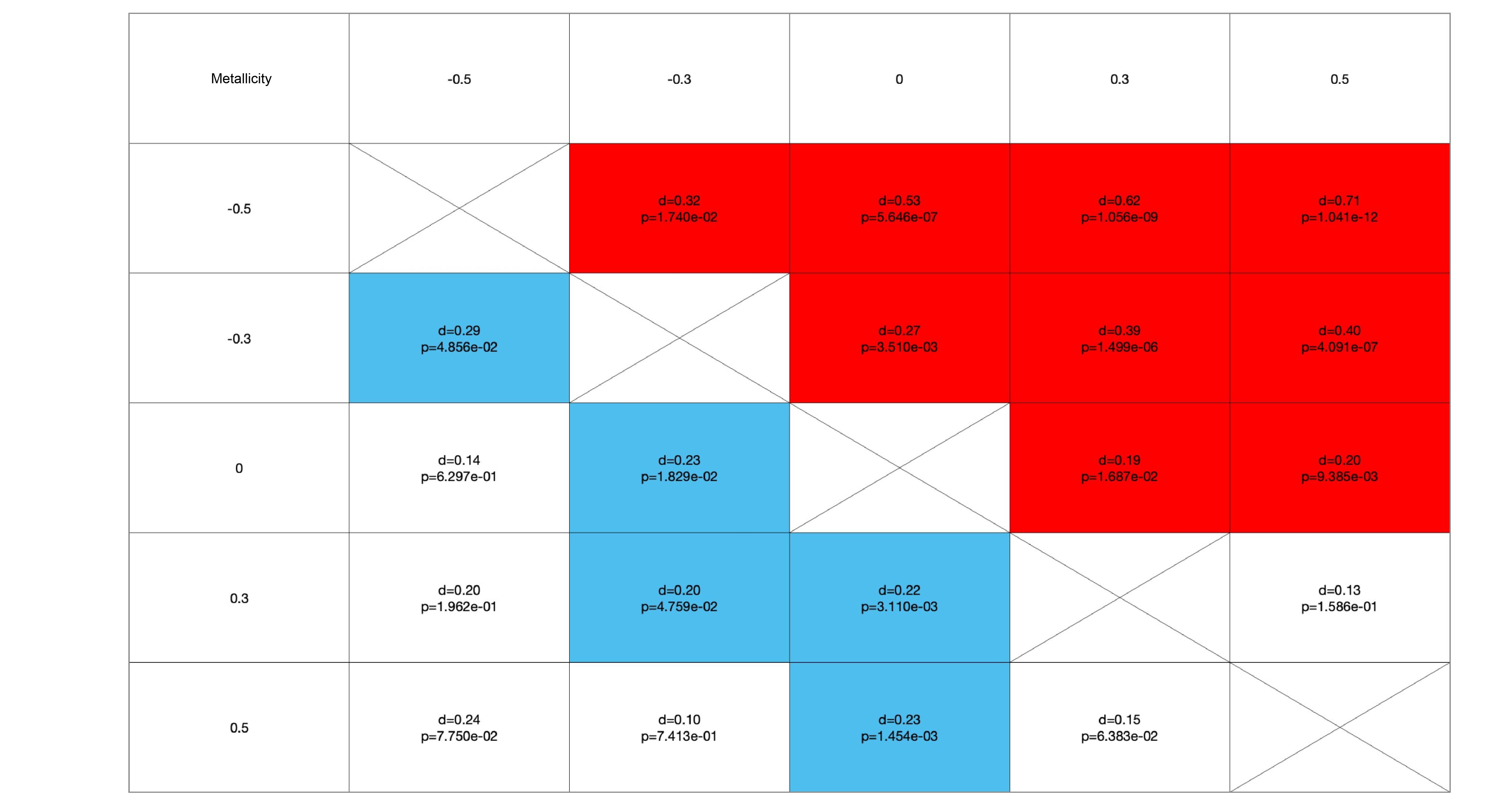}
\caption{KS test results for metallicity analysis with \(\lambda = 3\) for high-mass planets. 
The bottom-left triangle represents the test results for planetary mass, while the top-right triangle represents the results for orbital radius. 
The two-sample KS statistic $d$ and the corresponding probability $p$ are shown for each metallicity sample pair.
Grids with p-values less than 0.05 for orbital radius are highlighted in red, and those for planetary masses are highlighted in cyan.}
\label{fig:ks_test_gt30}
\end{figure}

\begin{figure}[h!]
\centering
\includegraphics[width=0.98\textwidth]{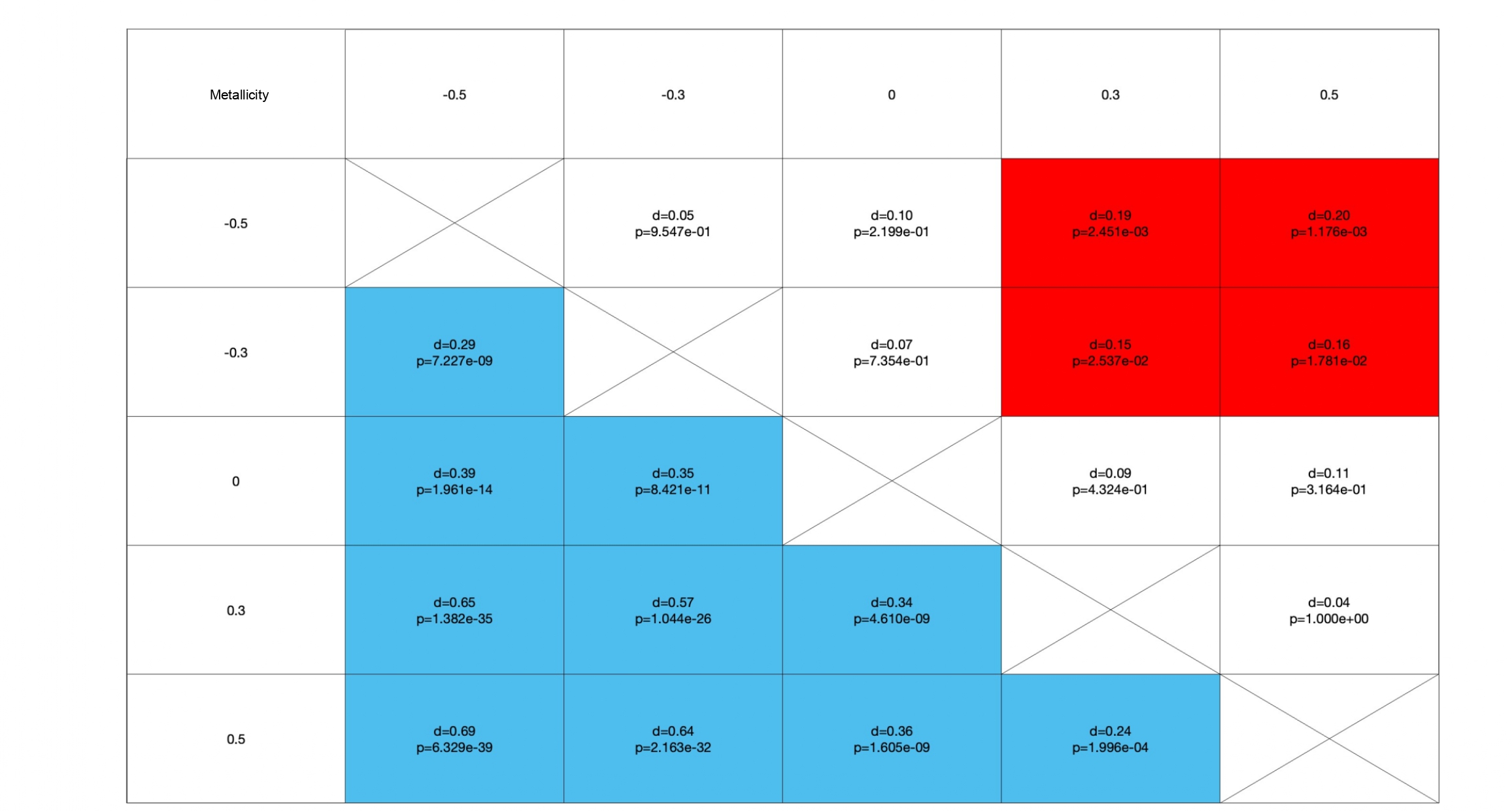}
\caption{Similar to Fig.~\ref{fig:ks_test_gt30} but showing the results of low-mass planets.}
\label{fig:ks_test_lt30}
\end{figure}

\twocolumn
\end{appendix}
\end{document}